# Where to place the positive muon in the Periodic Table?

Mohammad Goli and Shant Shahbazian[*]

*Faculty of Chemistry, Shahid Beheshti University, G. C. , Evin, Tehran, Iran, 19839, P.O. Box 19395-4716.*

Tel/Fax: 98-21-22431661

E-mail:
(Shant Shahbazian) chemist_shant@yahoo.com

[*] Corresponding author

# Abstract

In a recent study it was suggested that the positively charged muon is capable of forming its own "atoms in molecules" (AIM) in the muonic hydrogen-like molecules, composed of two electrons, a muon and one of the hydrogen's isotopes, thus deserved to be placed in the Periodic Table [Phys. Chem. Chem. Phys. 16, 6602, 2014]. In present report, the capacity of the positively charged muon in forming its own AIM is considered in a large set of molecules replacing muons with all protons in the hydrides of the second and third rows of the Periodic Table. Accordingly, in a comparative study the wavefunctions of both sets of hydrides and their muonic congeners are first derived beyond the Born-Oppenheimer (BO) paradigm, assuming protons and muons as quantum waves instead of clamped particles. Then, the non-BO wavefunctions are used to derive the AIM structures of both hydrides and muonic congeners within context of the multi-component quantum theory of atoms in molecules. The results of the analysis demonstrate that muons are generally capable of forming their own atomic basins and the properties of these basins are not fundamentally different from those AIM containing protons. Particularly, the bonding modes in the muonic species seem to be qualitatively similar to their congener hydrides and no new bonding models is required to describe the bonding of muons to a diverse set of neighboring atoms. All in all, the positively charged muon is similar to proton from structural and bonding viewpoint and deserved to be placed in the same box of hydrogen in the Periodic Table. This conclusion is in line with a large body of studies on the chemical kinetics of the muonic molecules portraying the positively charged muon as a lighter isotope of hydrogen.

# 1. Introduction

In recent decades there has been a growing interest in the chemistry of the exotic species, i.e. atomic and molecular species containing fundamental particles other than electrons and usual nuclei, where the muonic molecules are iconic examples.[1-5] Muons, as members of the lepton family, are heavier congeners of electrons, $\sim 206.8 m_e$; they are fermions appeared in both positively and negatively charged versions with a micro-second life time, $\sim 2.2 \times 10^{-6}$ s.[6] The negatively charged muons, $\mu^-$, are particularly well-known for their predominant role in the muon catalyzed "cold fusion" that once was considered as a serious alternative to the fission based nuclear technology.[3,4] In this process $\mu^-$, instead of electrons, acts as a "glue" bringing two hydrogen isotopes in a hydrogen molecule into very close contact elevating the probability of the nuclear fusion considerably.[3] Alternatively, in the bound states of muonic molecules $\mu^-$ encircles one of the nuclei in very tight orbits effectively reducing/screening one unit of the atomic number yielding a muon-induced "transmutation".[7-9] The positively charged muon, $\mu^+$, is usually assumed as a lighter isotope of hydrogen,[10,11] and this has been strengthened further with recent comparative chemical kinetics studies.[12-16] $\mu^+$ is also used as a probe in various muon spin spectroscopes that recently have found vast applications in the field of radical chemistry and beyond.[17-29] In a recent computational study we have proposed that $\mu^+$ resembles a light isotope of hydrogen not only from kinetics view but also from "structural" viewpoint since it is capable of forming its own "atomic basin" within the muonic hydrogen molecule.[30] In other words, $\mu^+$ is competent of accumulating electrons, forming a muon-electron(s) "cluster" that acts as an *atom in a molecule*.[31] This ability of $\mu^+$ was uncovered through analyzing the ab initio wavefunctions, derived for various congeners of hydrogen molecule, where one of the orthodox isotopes of hydrogen was substituted with $\mu^+$. The analysis done by the newly developed Multi-Component Quantum Theory of Atoms In Molecules (MC-QTAIM) method,[32-39] indicated that $\mu^+$ in competition with the heavier hydrogen nuclei has a lesser capacity to maintain electrons in its own atomic basin.[30] Accordingly, the question emerges that $\mu^+$ in competition with the nuclei of heavier elements still is capable of forming its own atomic basin. Present study is a primary attempt to answer this question.

The Periodic Table (PT) has been organized using the trends observed in various basic macroscopic, e.g. melting and boiling temperatures, mass densities, etc., and "microscopic", e.g. oxidation numbers, electronegativity, atomic volumes and radii, etc., properties of the chemical elements. While chemists introduce atoms as the basic microscopic building blocks and the pillar of introducing microscopic properties of elements, what is really in back of their mind is an atom/element implanted in a "chemical environment", e.g. a molecule or a crystal, not just "free" atoms.[40] As a result, most of microscopic properties are describing aspects of "atom's response" to its environment, e.g. oxidations states or electronegativity, where the theoretical background for such properties is the concept of "atom in molecule/crystal". This concept is the basic unit that not only reveals the identity and properties of an element at microscopic level, but also reveals its slight variations in response to various environments. Even the macroscopic/bulk properties of elements are eventually explained and recognized according to the nature of mutual interaction of atoms of an element thus inevitably tied to the properties of "atom in bulk". *Accordingly, the PT is not the collection of the free atoms but basically the organization of atoms in molecules (AIM) based on their properties.* Each atom in a molecule is a microscopic representative of an element in a chemical environment and the "box" of an element in the PT contains the properties of various typical AIM of an element, e.g. AIM with various oxidation numbers and atomic volumes. Although the concept of AIM has its root in nineteenth century chemistry, deep in the heart of the Structural theory of chemistry,[41] the modern incarnation of the AIM is within the context of the QTAIM.[42-44] The QTAIM methodology aims to extract real-space picture of a molecule or a crystal from its corresponding wavefunction deduced from Schrödinger's equation thus making a bridge between chemistry and quantum mechanics. Accordingly, the ab initio derived wavefunction is the "input" of the QTAIM analysis and the AIM morphology and properties are the "output". However, the orthodox QTAIM is only capable of dealing with electronic wavefunctions where just electrons are treated as quantum waves but the nuclei as clamped particles. In other words, the orthodox QTAIM is inherently a "single-component" methodology unable to deal with systems containing two or more "types" of quantum waves. Even in the case of usual molecular species this is a shortcoming since beyond the Born-Oppenheimer (BO) approximation and the concomitant clamped nuclei model, molecules are "multi-component" quantum systems and each nucleus must be treated as a quantum wave instead of a clamped particle. In the case of exotic species, the orthodox QTAIM is not applicable even within the BO approximation since both electrons and the light exotic particles of the molecular system under study must be treated as quantum waves. Thus, the exotic species are genuinely multi-component systems and their AIM structure remains elusive from the standpoint of the single-component theory. To overcome these deficiencies and extending the AIM analysis to the ab initio multi-component wavefunctions we have recently introduced an extended QTAIM methodology termed MC-QTAIM.[30,32-39] The MC-QTAIM methodology use both single-component and multi-component wavefunctions as input delivering the AIM morphology and their properties. The results of the MC-QTAIM analysis are indistinguishable from the orthodox QTAIM when the masses of all the particles that are treated as quantum waves, except electrons, tends to infinity demonstrating the fact that MC-QTAIM formalism encompasses the QTAIM as an "asymptote".[34] This observation clearly points to the fact that the MC-QTAIM is a unified scheme revealing the AIM structure of a large and traditionally unrelated sets of molecular systems.

To cope with the question of how incorporating $\mu^+$ into the PT, we hypothesize that *a fundamental or a combined exotic particle deserves to be a member of the PT, having an independent chemical identity, if it is in general able to form its own AIM, composed only from the exotic particle and electrons, when combined with the ordinary matter forming bound states*. Based on this supposition, positron, the anti-particle of electron, does not deserved to be placed in the PT since previous computational studies revealed that this particle is unable to form independent AIM.[39,45-47] On the other hand, the muonic Helium as a combined exotic particle, composed of an alpha particle and an encircling $\mu^-$, deserves to be placed in the same box of hydrogen in the PT since it forms its own AIM quite similar to proton containing atomic basins.[30] The short life time of the muonic Helium composite is not an obstacle since many radioactive and superheavy nuclei, composed of protons and neutrons, may have also exceedingly short lives but still to be placed in the PT.[48-50] Based on this reasoning, $\mu^+$ ability to form AIM in various chemical environments is what must be checked to determine whether it deserves to be placed in the PT. On the other hand, the similarity or dissimilarity of these $\mu^+$ containing AIM to those of proton containing AIM is a gauge to judge whether $\mu^+$ must be placed in the box of hydrogen (as a member of hydrogen family), or independently in a new box for instance before hydrogen's box. In this study as a primary test hydrides of the second and third row elements of the PT a re considered from the standpoint of their AIM structure replacing protons with $\mu^+$.

## 2. Computational Details

In order to perform the MC-QTAIM analysis, the multi-component wavefunctions of each species were derived employing the Nuclear-Electronic Orbital (NEO) methodology to perform ab initio calculations on $MX_k$ species including the series LiX, $BeX_2$, $BX_3$, $CX_4$, $NX_3$, $OX_2$, FX from the second row and the series NaX, $MgX_2$, $AlX_3$, $SiX_4$, $PX_3$, $SX_2$, ClX from the third row of the PT where X stands for proton (H) or $\mu$ (in chemical formulas $\mu^+$ is hereafter abbreviated as $\mu$). The NEO ab initio methodology, which has been developed by Hammes-Schiffer and coworkers,[51] tries to solve Schrödinger's equations for the multi-component systems. The general strategy of the NEO is based on a hierarchical structure starting from the mean-field approximation and then proceeding further by employing more complex wavefunctions trying to simulate more closely the exact but unattainable multi-component wavefunction. Briefly, at first stage the multi-component Hartree-Fock (HF) equations are solved as the first-order approximation and then the configuration interaction (CI) method is used to improve the original multi-component HF wavefunction; the latter is a product of Slater determinants, each describing one type of quantum particles. The CI wavefunction is an expansion of electron-proton or electron-muon configurations where each is the product of the Slater determinants constructed from the virtual orbitals as the byproducts of the algebraic solution of the multi-component HF equations. The resulting methods are termed NEO-HF and NEO-CI while the details of the theory may be found elsewhere;[51] the original NEO computer code, containing both the NEO-HF and NEO-CI methods, has been implemented into the GAMESS suite of programs.[52] We have modified the original code adding new capabilities making it suitable for ab initio calculations on $\mu^+$ containing species,[38,39] as also have been done previously by others.[53]

To perform the ab initio calculations protons, muons and electrons in all the species were equally treated as quantum waves and the used approximate masses for proton and $\mu^+$ are $1836m_e$ and $206m_e$, respectively, ($m_e$ stands for electron's mass), while the heavier nuclei were treated as clamped point charges. Throughout calculations the spin states of electrons were assumed to be singlet closed-shell while for protons and muons all individual spins directions were assumed to be parallel yielding a total high-spin open-shell multiplet state. The standard 6-311+g(d) basis set was used for the clamped nuclei to expand the electronic orbitals.[54] For each quantum proton and $\mu^+$ a [4s1p] combination was used as the electronic basis set and the exponents of all basis functions were optimized variationally for each species. This strategy was employed to ensure that the electronic basis set is flexible enough to respond properly to the slight changes induced by replacing proton with $\mu^+$. A single s-type Gaussian basis function was used to represent the one-particle wavefunction of each quantum proton/$\mu^+$ and the exponents were optimized variationally. A single joint center, represented by a Banquet/ghost atom and abbreviated hereafter as Bq, was located for both electronic and protonic/muonic basis sets. The geometries of the optimized clamped hydrides at HF/6-311++g(d,p) level were used as an initial guess replacing each clamped hydrogen nucleus with a Bq. The original point group symmetries, $D_{\infty h}$ ($BeX_2$ and $MgX_2$), $D_{3h}$ ($BX_3$ and $AlX_3$), $T_d$ ($CX_4$ and $SiX_4$), $C_{3v}$ ($NX_3$ and $PX_3$), $C_{2v}$ ($OX_2$ and $SX_2$), were preserved throughout calculations for the "pseudo-geometries" where Bqs now used to define point groups. Next, for each species the position of Bqs and also the exponents of basis functions (except from those basis functions located at the clamped nucleus) and the self-consistent field (SCF) coefficients of all the basis functions were optimized simultaneously during the NEO-HF calculations. The iterations of the SCF cycles were terminated when the total NEO-HF energy variations were below $10^{-10}$ Hartree. To ensure the quality of the non-linear optimization procedure, the gradients of energy, i.e. forces, were computed on both the clamped nuclei and Bqs and it was demonstrated that in all of the considered species the final computed root mean square (RMS) and maximum of energy gradients were always below $10^{-5}$ Hartree/Bohr.

At next stage, to study the role of electron correlation, a truncated NEO-CI calculation was done for each species starting from the NEO-HF/[6-311+g(d)/4s1p:1s] wavefunction. To select the electronic configurations an "active space", denoted as ($n,m$) pair, where $n$ stands for the active electrons and $m$ for the active orbitals, was selected for each species and all possible Slater determinants were produced from this active space using virtual orbitals derived from the previous NEO-HF calculation. Subsequently, the NEO-CI expansion was constructed from the joint product of the muonic NEO-HF determinant and the electronic determinants derived from the active space while the variational optimization of the CI expansion coefficients yields the final optimized CI wavefunction. No exponent or center optimization was done at this stage thus the MC-QTAIM analysis of the resulting NEO-CI/[6-311+g(d)/4s1p:1s] wavefunction just reveals the pure role of the electron correlation. The wavefunctions produced in the above mentioned ab initio procedures were then used for the MC-QTAIM analysis.

To begin the MC-QTAIM analysis each multi-component wavefunction was transformed to an "extended" WFN protocol involving the positions of the clamped nuclei and Bqs as well as the mass, charge and statistics of the particles treated as quantum waves. In the case of the NEO-HF wavefunctions both the optimized canonical spatial orbitals of electronic and muonic determinants were implemented in the extended protocol while in the case of the NEO-CI wavefunctions the natural electronic spatial orbitals were used instead. The whole procedure has been automated and incorporated into the modified NEO code and will be disclosed in detail in a future publication. The algorithm used for the MC-QTAIM analysis has been described comprehensively elsewhere and its machinery is not reiterated here.[32,46] The MC-QTAIM analysis was done first performing the topological analysis of the Gamma density (*vide infra*) and deciphering the boundary of AIM, i.e. inter-atomic surfaces.[42-44] In next stage, the combined property densities were integrated in each atomic basin yielding the atomic properties. Since the computed ab initio virial ratio, $\langle V \rangle / \langle T \rangle$, for some species deviate from its exact value, $-2$ (see Table 1), an *ad hoc* virial scaling were done in computing atomic energies.[46] Checking the accuracy of the numerical basin integration algorithm, the net flux integral, $\widetilde{L}(\Omega) = (-1/4) \int_\Omega d\vec{q} \ \nabla^2 \Gamma^{(2)}(\vec{q})$, was computed for each atomic basin as a standard gauge and in all cases it was demonstrated that $L(\Omega) < 10^{-4}$ (in atomic units). As a final test of the quality of basin integrations, the sum of the MC-QTAIM derived atomic properties, e.g. basin energies, was compared with those derived independently for molecules from the ab initio calculations and the comparison demonstrated the precision of the numerical basin integration procedures. The figures displaying Molecular Graphs (MGs) and atomic boundaries were all constructed using the AIMALL package.[55] Throughout the paper all results are offered in atomic units (au).

## 3. Ab initio calculations

Table 1 and Figure 1 offer selected results of the ab initio calculations at the NEO-HF/[6-311+g(d)/4s1p:1s] computational level while in discussing the emergent patterns it is always assumed throughout the paper that all trends for each series of species are described from the molecule containing lightest to the one containing the heaviest central atom thus the phrase "scanning the row from the left to the right-side of the PT" is eliminated from corresponding statements. For comparison of the pseudo-geometries of the hydride/protonic and muonic species, "mean" inter-nuclear ($M-X$) distances and angles ($X-M-X$) are introduced as the distance between the central atom and each of Bqs and the angle between two Bqs and the central atom, respectively. By employing [1s] protonic/muonic basis set the distribution of each proton or $\mu^+$ is described by a single Gaussian function centered at a Bq and the computed mean inter-nuclear distances are also the expectation values of the $M-X$ separation operator. In each series of the protonic species the mean $M-H$ distances diminish, as is expected,[56] and interestingly, the same pattern also emerges considering the mean $M-\mu^+$ distances in both of the muonic series of species. Figure 1a depicts the difference between the mean inter-nuclear distances of each pair of the congener protonic and muonic species, e. g. $MH_3$ versus $M\mu_3$, revealing the fact that the $M-\mu^+$ distances are almost constantly, $0.10\pm0.2$ au, longer than the $M-H$ distances. In a previous study on the muonic hydrogen molecules it was also observed that upon replacing one of protons of the usual hydrogen molecule with $\mu^+$, the mean inter-nuclear distance increases, ~0.1 au.[30] No such variation is observable in $X-M-X$ and one may claim that replacing $\mu^+$ with proton has virtually no effect on the mean angles. The total kinetic energy of protons/muons, $K_X$, in each species has been computed and presented in Table 1, however, the kinetic energy depends on the number of particles. Thus, for a more clear picture Figure 1b presents the kinetic energy per particle, $k_x = K_X/N_X$ ($N_X$ stands for the number of protons/muons of each species), for both protonic and muonic species. Evidently, in contrast to $K_X$, $k_x$ is almost constant for the protonic, $k_H = 0.018 \pm 0.002$ au, and the muonic species, $k_\mu = 0.041 \pm 0.004$ au, and considerably larger for the latter species. Upon attributing a single s-type Gaussian one-particle wavefunction to each proton/$\mu^+$, it is straightforward to demonstrate *analytically* that $k_x = 3\alpha_X/2m_X$ where $\alpha_X$ and $m_X$ stand for the exponent of the Gaussian function and the mass of particle, respectively;[34] Figures 1c and 1d depict $k_x$ versus variationally optimized $\alpha_X$ in both protonic and muonic series revealing the expected linear graph. Since the variations of the computed $\alpha_X$ in both protonic and muonic species are in a small range, $\alpha_H = 21.5 \pm 2.0$ and $\alpha_\mu = 5.6 \pm 0.5$, the concomitant variations of $k_x$ are also confined to a small range. The variationally derived exponents unravel the fact that they are mainly influenced by the *identity* of the particles and their ratio is almost constant for each pair of the congener protonic and muonic species, $\alpha_\mu/\alpha_H \approx 0.26$; "chemical environment" only marginally affects their values. Clearly, muonic one-particle wavefunctions (orbitals) are more spatially extended than those of protons and roughly speaking, muons are vibrating with larger amplitudes than protons.

To have a more detailed picture one may use the fact that the s-type Gaussian function is also the ground state eigenfunction of the 3D isotropic harmonic oscillator.[34] It is straightforward to demonstrate *analytically* that the force constant and zero-point energy (ZPE) of this oscillator are given by $f_X = 4\alpha_X^2/m_X$ and $E_{ZPE}^X = 3\alpha_X/m_X = 2k_x$, respectively. In comparison with each congener pair of the protonic and muonic species it emerges that $f_\mu/f_H = 0.62 \pm 0.02$ and $E_{ZPE}^\mu/E_{ZPE}^H = 2.33 \pm 0.03$. Accordingly, it seems that upon replacing $\mu^+$ with proton in a molecule one is faced with more energetic and less stiff vibrations with larger amplitudes. This picture is also in line with the previously mentioned elongation of the mean $M-\mu^+$ distances relative to the mean $M-H$ distances pointing to "expansion" of the "pseudo-nuclear" framework of the muonic species in comparison to their protonic congeners. Through the SCF procedure of the NEO-HF, the more extended muonic orbitals are less capable of accumulating electrons around themselves than the protonic orbitals thus the electronic distribution also expands relative to the protonic congeners. Although the electronic expansion is best manifested in the MC-QTAIM analysis (*vide infra*), even at this stage it is traceable in ab initio data. Comparing total energies of the congener pair of the protonic and muonic species reveals that the total energy of the protonic species is always more negative than their muonic counterparts. Through the virial theorem, $E_{total} = -K_e - K_X$ ($K_e$ stands for the electronic kinetic energy), one inevitably comes to the conclusion that the electronic kinetic energies of the protonic species are larger than their muonic congeners in line with the computed electronic kinetic energies in Table 1. Manifestly, the expanded electronic distribution of the muonic species yields a smaller electronic kinetic energy relative to their protonic congeners, which dominates the total energy, in contrast to the fact that protons themselves have smaller kinetic energies than muons. As a final check, Table S1 in supporting information offers the variationally optimized exponents of the electronic basis functions for all the considered species. Clearly, the mean of the exponents in the protonic species, $8.4 \pm 2.2$, $1.9 \pm 0.6$, $0.5 \pm 0.2$, $0.2 \pm 0.1$ for the s-type Gaussian functions and $0.7 \pm 0.4$ for the p-type Gaussian function, are larger than those of the muonic species, $4.4 \pm 1.2$, $1.3 \pm 0.5$, $0.4 \pm 0.1$, $0.1 \pm 0.1$ for the s-type Gaussian functions and $0.6 \pm 0.4$ for the p-type Gaussian function. One may conclude that in comparison of the protonic and muonic congeners both pseudo-nuclear framework and electronic distribution of the muonic species expand concomitantly.

Table S2 in supporting information offers selected results of the ab initio calculations at the NEO-CI/[6-311+g(d)/4s1p:1s] computational level. The computed electronic correlation energies are small and no serious deviations are observable from the computed results at the NEO-HF/[6-311+g(d)/4s1p:1s] level. Therefore, one may conclude that the inclusion of the electronic correlation does not change the previous discussions qualitatively. As will be discussed in next section, the same conclusion emerges from considering the topological analysis at NEO-HF and NEO-CI levels.

## 4. The MC-QTAIM analysis

### 4.1. Topological analysis

The MC-QTAIM analysis begins with the topological analysis of the Gamma density, $\Gamma^{(2)}(\vec{q}) = \rho_e(\vec{q}) + (1/m_X)\rho_X(\vec{q})$ (written in atomic units), where $\rho_e(\vec{q}) = N_e \int d\tau'_e \Psi^* \Psi$ is the one-particle density of electrons while $\rho_X(\vec{q}) = N_X \int d\tau'_X \Psi^* \Psi$ is the one-particle density of protons/muons.[35] In these equations $N_e$ and $\Psi$ are the number of electrons and the multi-component wavefunction while $d\tau'_t$, $t = e, X$ implies summing over spin variables of all quantum particles and integrating over spatial coordinates of all quantum particles except one arbitrary particle belonging to the subset of electrons, denoted by subscript $e$, or the subset of protons/muons, denoted by subscript $X$. The critical points (CPs) are determined through constructing the gradient vector field of the Gamma density, $\vec{\nabla}\Gamma^{(2)}(\vec{q})$, and searching for the attractors of the field, $\vec{\nabla}\Gamma^{(2)}(\vec{q}_{CP}) = 0$. In next stage, selected "property densities" are computed at CPs, termed topological indices, where each property density is a combined density originating from both electrons and protons/muons, $\tilde{M}(\vec{q}) = M_e(\vec{q}) + M_X(\vec{q})$. At first the qualitative results of the topological analysis are discussed and then quantitative aspects are considered to have a more detailed picture of the similarities and dissimilarities in bonding pattern of the protonic and muonic series. For quantitative considerations Tables 2 and 3 offer selected topological indices, $\widetilde{M}(\vec{q})$, computed at the (3, -1) CPs and also at the (3, -3) CPs located at (or very near to) the Bqs while Tables 4, S3 and S4 (the latter two in supporting information) offer the separate electronic, $M_e(\vec{q})$, and protonic/muonic, $M_X(\vec{q})$, contributions.

In the case of the protonic species the topological analysis reveals MGs that are topologically equivalent to those derived within the context of the orthodox QTAIM except FH molecule (see Figure S1 in supporting information). In the all protonic species at (or very near to) each Bq a (3, -3) CP emerges linked to the (3, -3) CP centered at the central atom with a (3, -1) CP in between; based on our recent proposal, to circumvent any misinterpretation, hereafter a (3, -1) CP is called a "line" critical point (LCP) instead of "bond" critical point.[57,58] The only exception is FH molecule where just a single (3, -3) CP emerges very near to the clamped fluorine nucleus with no (3, -3) CP near the Bq. Even within the context of the orthodox QTAIM, at the HF/6-311++g(d,p) level, the (3, -3) CP at hydrogen nucleus is too close to the LCP (~0.2 au) and the amount of one-electron density at the (3, -3) CP, ~0.42 au, is very near to that at the LCP, ~0.40 au. Interestingly, the ionized FH molecule, $FH^+$, is also one of the rare examples that (3, -3) CP on the clamped proton may disappear during variations of the inter-nuclear distance.[59] Therefore, it is quite probable that any small variation of the Gamma density around the Bq, originating from the usage of a more advanced computational level or a more rigorous ab initio non-BO methodology, may yield the "missed" (3, -3) CP and the LCP in between (3, -3) CPs. We leave this possibility as an open problem for future studies since in a more rigorous ab initio study both nuclei, not just the proton, must be treated as quantum waves and the translational and rotational motions must be explicitly taken into account and excluded from the multi-component wavefunction.[60] The MGs of all the muonic species, except $O\mu_2$, are topologically equivalent to those of their protonic congeners and have been depicted in Figures 2 and 3 for the species containing the central atoms from the second row of the PT (see also Figures S2 in supporting information). Interestingly, in a previous study it was demonstrated that while $\mu^+$ is capable of forming its own atomic basin replacing proton in $H-C\equiv N$ species, it is unable to form a (3, -3) CP in $C=N-H$ species when replacing proton.[39] In the latter case, as well as in $F\mu$ and $O\mu_2$ species, $\mu^+$ is "competing" with a highly electronegative atom on its share of electrons and is not capable of accumulating enough electrons around the muonic distribution to shape an attractor in the gradient vector field of the Gamma density.[31] As has been proposed recently,[39] in all such cases that $\mu^+$ is unable to form its own atomic basin these species are better described as $(CN^-,\mu^+)$, $(O^{2-},2\mu^+)$ and $(F^-,\mu^+)$ rather than $CN\mu$, $O\mu_2$ and $F\mu$. However, like the case of FH, it cannot be excluded that in future more rigorous ab initio computational studies on these species new (3, -3) CPs may emerge around the Bqs associated to muons. Nevertheless, even at current computational level one may claim that $\mu^+$ *is capable of forming its own (3, -3) CP in competition with most elements of the second and third rows of the PT except probably from the most electronegative oxygen and fluorine atoms*. The recently proposed topological floppiness index,[36] $TF = \Gamma(\vec{q}_{(3,-1)CP})/\Gamma(\vec{q}_{(3,-3)CP-X})$ (X stands for proton/$\mu^+$) quantifies this picture. In all the four series of species the $TF$ index increases from $0.15\pm0.03$ toward its limiting value, $TF=1$, while a comparison of the congener protonic and muonic species demonstrates that the $TF$ is always larger for the muonic congener. Evidently, in combination with a certain atom from the second or third row of the PT, compared to proton, $\mu^+$ is less capable of accumulating the Gamma density around itself and the resulting topological structures, i.e. MGs, are more prone to variations because of external perturbations.[36] In other words, muonic species have a more "floppy" topological structures than their protonic congeners and this is best seen particularly in the series of the second row species where the *TF* is very near to its limiting value for $N\mu_3$, ~ 0.987; while proton is unable to form a (3, -3) CP beyond oxygen atom, $\mu^+$ lacks that capability beyond nitrogen atom. This is also in line with the computed lengths of the line paths, i.e. gradient paths, connecting LCPs to (3, -3) CPs at or near to Bqs, which decrease intensely in all the four series of species.

To have a more precise picture selected topological indices were computed at both LCPs and (3, -3) CPs associated with Bqs including the Laplacian of the Gamma density, $\nabla^2 \Gamma^{(2)}(\vec{q}_{CP}) = \nabla^2 \rho_e(\vec{q}_{CP}) + (1/m_X)\nabla^2 \rho_X(\vec{q}_{CP})$, the combined Lagrangian kinetic energy density, $\widetilde{G}(\vec{q}_{CP}) = G_e(\vec{q}_{CP}) + G_X(\vec{q}_{CP})$, and the combined Hamiltonian energy density, $\widetilde{H}(\vec{q}_{CP}) = H_e(\vec{q}_{CP}) + H_X(\vec{q}_{CP})$, the latter just at LCPs, all introduced and detailed previously.[34,35] Tables 4, S3 and S4 reveal the origin of observed patterns in the combined densities by offering the separate electronic and protonic/muonic contributions. Evidently, at the LCPs of the protonic species, except H$_2$O molecule (and to a much lesser extent NH$_3$), the protonic contribution is essentially null and the combined densities originate virtually from electrons, $\widetilde{M}(\vec{q}_{LCP}) = M_e(\vec{q}_{LCP})$. Interestingly, and in contrast to the picture emerging from the orthodox QTAIM,[42] in the case of H$_2$O molecule protons' contribution to the topological indices at the LCPs is not null pointing to the fact that each proton "leaks" slightly into the oxygen basin. The computed proton population, offered in Table 5, conforms to this picture and demonstrates that in H$_2$O molecule the proton population is not fully contained within the hydrogen basin, $N_H(\Omega_H) \approx 0.991$ ($N_H(\Omega_O) \approx 0.009$). For the muonic species, the leakage of $\mu^+$ is observed mainly for $N\mu_3$ and $Cl\mu$ species (and to lesser extent in $C\mu_4$ and $S\mu_2$) and the computed $\mu^+$ populations in atomic basins, offered in Table 6, conform to this picture; the $\mu^+$ leakage in $N\mu_3$, $N_\mu(\Omega_\mu) \approx 0.944$ ($N_\mu(\Omega_N) \approx 0.056$), and $Cl\mu$, $N_\mu(\Omega_\mu) \approx 0.986$ ($N_\mu(\Omega_{Cl}) \approx 0.014$), are larger than that of the protons' in H₂O molecule. Indeed, according to the previously derived ab initio results muons have larger vibrational amplitudes than protons thus more prone to leakage into neighboring atomic basin. Thus, while one needs to electronegative oxygen atom as a neighbor to observe the proton leakage from hydrogen basins, less electronegative atoms as neighbors suffice to observe the leakage of $\mu^+$ from its own basin. The dissection of the Gamma and combined densities at the (3, -3) CPs associated to Bqs complements the picture emerged from the same analysis at LCPs. Comparison of Tables 4 and S4 reveals that the contribution of the electronic one-particle density is one order of magnitude larger than the mass-scaled one-particle density of protons/muons contributing to the Gamma density, $(1/m_X)\rho_X(\vec{q})$, which is almost constant, $0.030 \pm 0.005$, in all the four series of species. As a result, the variation of the Gamma density at the (3, -3) CPs in each of the series of species is practically dictated by the electronic one-particle density. This is not the case when considering the Laplacian of the Gamma density and the Lagrangian kinetic energy density where the electronic and protonic/muonic contributions are within the same order of magnitude and almost comparable. The comparison of the non-scaled one-particle protonic and muonic densities as well as their Laplacians at the (3, -3) CPs associated to Bqs indeed confirms that muons are less localized than protons. This sparseness of the muonic one-particle density also reflects itself in smaller electronic one-particle density as well as larger (less negative) corresponding Laplacians at (3, -3) CPs associated to Bqs in the muonic species in comparison to their protonic congeners. These results are compatible with previous observations and point beyond any doubt to the fact that the muonic one-particle densities, through the SCF procedure of the NEO-HF method, are less capable of accumulating electrons around themselves than the more localized protonic one-particle densities.

To have a comparative picture of bonding modes in the protonic and muonic species, the electronic contribution of selected topological indices are considered which have been used within context of the orthodox QTAIM as indicators of bonding.[42,43,61,62] These include $\rho_e(\vec{q}_{LCP})$, $\nabla^2\rho_e(\vec{q}_{LCP})$, $G_e(\vec{q}_{LCP})$, $H_e(\vec{q}_{LCP})$, $G_e(\vec{q}_{LCP})/\rho_e(\vec{q}_{LCP})$ which are routinely used to distinguish ionic and covalent bonding modes in molecules composed of the main group elements. The general patterns of variations of all the topological indices in the two protonic series indicate that in each series the ionic contribution to the bonding decreases while the covalent character increases. Accordingly, the one-particle density of electrons, the corresponding Laplacian, the electronic Hamiltonian energy density and the ratio of the electronic Lagrangian kinetic energy density to the electronic one-particle density all decrease. This is also generally in line with the usual orbital-based models used to rationalize the bonding modes of the main group hydrides,[56] though a comprehensive and detailed view of bonding only emerges after considering the results of basin integrations (*vide infra*). Exactly the same trends are also observable from the computed topological indices in the two muonic series and just small quantitative differences between the topological indices of the congener protonic and muonic species are observable. Nevertheless, such subtle differences do not seem to force one a radical departure from the bonding schemes conceived for the congener protonic species and at least in a qualitative viewpoint muon's bonding capability seems similar to that of the proton's.

Finally, it is illustrative to compare the topological analysis done on the two muonic series using the NEO-CI/[6-311+g(d)/4s1p:1s] wavefunctions, offered in Table S5 in supporting information, with that performed at the NEO-HF/[6-311+g(d)/4s1p:1s] level. Comparison of Tables 2 and 3 with Table S5 unravels that topological analysis not only qualitatively, i.e. the number and types of CPs, but also quantitatively, i.e. topological indices computed at CPs, is not sensitive to electron correlation thus the results and patterns observed at NEO-HF level are virtually unaltered at NEO-CI level. Consequently, because of the observed insensitivity of the topological analysis to the inclusion of the electron correlation all subsequent discussions are confined to the NEO-HF wavefunctions.

### 4.2. Properties of AIM

To perform the basin integrations, the inter-atomic surfaces are first derived from the local zero-flux equation of the Gamma density, $\vec{\nabla}\Gamma^{(2)}(\vec{q})\cdot\vec{n}(\vec{q})=0$, as zero-flux surfaces each going through one of LCPs.[35] The LCP laying on the inter-atomic surface acts as the global attractor of all the gradient paths that are on the surface while the atomic basins are delineated by the inter-atomic surfaces each containing a (3, -3) CP that acts as the global attractor of all the gradient paths that are within the atomic basin.[42,43] In next stage, property densities are integrated within each atomic basin and the atomic properties are derived originating from both electrons and protons/muons, $\widetilde{M}(\Omega)=\int_{\Omega}d\vec{q}\ \widetilde{M}(\vec{q})=M_e(\Omega)+M_X(\Omega)$; the sum of atomic properties yields molecular property, $\sum_{\Omega}\widetilde{M}(\Omega)=M_{molecule}$. Selected results of the basin integrations are offered for the protonic and muonic species in Tables 5 and 6, respectively.

Based on the topological analysis discussed in previous subsection in all species each proton/$\mu^+$ has its own atomic basin except ($F^-,H^+$), ($O^{2-},2\mu^+$) and ($F^-,\mu^+$) which have just a single basin similar to free atoms. However, because of the protonic/muonic localized distribution around Bqs, in contrast to free atoms, the electronic one-particle density of these single-atom basins are not spherically distributed, e.g. one-electron density of a free atom in an external field. Since the focus of the present study is on the protonic/muonic basins, these three species are discarded from subsequent analyzes and discussions. The electronic populations of the protonic/muonic basins, $N_X^e(\Omega)$, in all the four series of species decrease in line with the well-known fact that the electronegativity of central atoms increases toward the right-hand side of the PT. The atomic charges of the congener protonic and muonic basins, $q_X(\Omega_X) = N_X(\Omega_X) - N_X^e(\Omega_X)$, offered in Figure 4a, are almost the same when the central atom is one of the first two elements of each row. But, for remaining elements as central atoms, the muonic basins have larger atomic charges in contrast to their protonic congeners. In a previous paper, based on "direct" competition of $\mu^+$ with various hydrogen isotopes on its share of electrons in hydrogen-like muonic molecules, it was proposed that $\mu^+$ has a smaller electronegativity, though not explicitly determined, than proton and heavier hydrogen isotopes.[30] The observed trend in the computed atomic charges in current study also conforms to the stated conclusion and widens it applicability domain to "indirect" competition; when proton and $\mu^+$ compete on their share of electrons with neighboring basin containing an element from the second or third row of the PT. However, as is also evident from Figure 4a, the different capacity of $\mu^+$ and proton in retaining electrons in their own atomic basins depends strongly on the nature of neighboring basin and is much more pronounced when the neighbor is a highly electronegative atom. Seemingly, even if one attributes a fixed electronegativity number to $\mu^+$, the charge transfer is not just a simple function of electronegativity difference and other factors, e.g. hardness and polarizability, are also probably important in the amount of the charge transfer.[63] This is an interesting problem that needs further theoretical and computational studies in future. In order to have a more detailed picture, the electron localization index ($LI^e$), which has been introduced recently for the multi-component wavefunctions within the context of the MC-QTAIM,[36] and the percent electron localization, $100 \times LI_X^e(\Omega)/N_X^e(\Omega)$, were computed and the latter offered in Figure 4b. Evidently, $LI_X^e$, similar to the pattern observed for the electronic population, decreases in all the four series of species. Also, it is larger for the protonic basins than their congener muonic basins except the species containing the first two elements of each row where practically no difference is observable. The computed percent electron localizations span a wide range, $\sim 95\%$ for LiH and $Li\mu$ molecules to $\sim 13\%$ for H$_2$O molecule, revealing the same patterns observed for the $LI$. As a result, all these further confirm the above mentioned conclusion that *in general $\mu^+$ has a lesser capacity than proton to retain electrons in its own atomic basin though this manifests itself unmistakably when $\mu^+$ and proton compete directly, or indirectly with an electronegative neighbor on their share of electrons.* Besides, the intra-atomic polarization dipoles of the atomic basins, just originating from electrons, were computed using Bqs as the centers of local coordinate system $\vec{P}_e(\Omega_X) = -\int_{\Omega_X} d\vec{q}\ (\vec{q} - \vec{R}_{Bq})\rho_e(\vec{q})$; since a single s-type Gaussian function has been attributed to each proton/$\mu^+$, one may demonstrate that the protonic/muonic contribution to the polarization dipole is null, $\vec{P}_X(\Omega_X) = 0$.[34,35] The computed dipoles demonstrate that muonic basins are generally more "deformed" than their protonic congeners revealing the fact that neighboring basins have a larger impact on the charge separation and polarization of the muonic rather than protonic basins.

An outer iso-density surface of the Gamma density encompassing the whole molecule is used to introduce the molecular volume while the atomic volumes are delineated by their inter-atomic surfaces and the iso-density surface.[36] The value of the Gamma density used for the outer surface is to some extent arbitrary though iso-density surfaces emerging from $\Gamma(\vec{q}) = 0.0004$, $0.001$, $0.002$ au equations are generally assumed to be good estimates of the molecular volume; in present study just $\Gamma(\vec{q}) = 0.001$ au is used to compute the atomic and the molecular volumes. Taking the fact that the iso-density surfaces of the protonic/muonic species are usually far from the concentration centers of the protonic/muonic one-particle densities, practically, the electronic one-particle density dictates the shape of the iso-density surfaces. Clearly, as is offered in Figure 4c, the molecular volumes of the muonic species are always larger than their protonic congeners in line with the electronic expansion of the muonic species inferred indirectly from the ab initio computed electronic kinetic energies and the optimized exponents of the electronic basis set. The atomic volumes of the protonic and muonic basins in all the four series of species decrease intensely also in line with the decrease of the length of the line paths connecting LCPs and the (3, -3) CPs associated to the protonic/muonic basins. A more detailed picture emerges comparing the difference in the atomic volumes of the muonic and the protonic basins of the congener species, $\Delta V = V(\Omega_\mu) - V(\Omega_H)$, offered also in Figure 4c. This comparison demonstrates that the difference, $\Delta V$, diminishes and for species containing central atoms from the right-hand side of the rows the volume of protonic basins are even larger than their muonic congeners.

Interestingly, this observation points to the fact that the muonic basins are *not* always larger than their congener protonic basins and the known volume expansion in the muonic molecules must be traced also in the concomitant expansion of the volume of the central atoms in the muonic species relative to their protonic congeners. Indeed in a recent study on the molecular volumes of selected organic molecules it was demonstrated that the "net" contraction of molecular volumes, induced by substitution of protons with deuterons, is the result of a "mixture" of volume expansions and contractions of constituent atomic basins.[38] *Inevitably, one must conclude that the expansion/contraction in the molecular volumes upon replacing protons with muons/deuterons is not always induced by the expansion/contraction of the muonic/deuteronic basins but the expansion/contraction of the neighboring basins must be taken also into account.* This interesting and seemingly unnoticed trend reveals a complex "compensatory" mechanism behind the variation of molecule volumes upon isotopic substitution that deserves further studies in future.

The basin energies are derived from the recently proposed local multi-component virial theorem,[34,35] which equates the basin energy to the basin integration of the minus sum of all the kinetic energy densities. Accordingly, for the protonic/muonic species one arrives at: $\widetilde{E}(\Omega) = -\int_\Omega d\vec{q}\{K_e(\vec{q}) + K_X(\vec{q})\} = -K_e(\Omega) - K_X(\Omega)$, where $K_X(\Omega)$ originates from protons/muons vibrations. However, for central atoms with no leakage of the protons/muons from neighboring basins, the basin energy just originates from the electronic kinetic energy, $\widetilde{E}(\Omega) = -K_e(\Omega)$. However, as was demonstrated recently,[38] when proton/$\mu^+$ distribution is effectively confined to its own basin, it is possible to derive two separate local virial theorems, one for electrons and another one for proton/$\mu^+$. In this case, the total basin energy is the sum of the separate electronic and protonic/muonic contributions: $\tilde{E}(\Omega) = E_e(\Omega) + E_X(\Omega)$ where $E_e(\Omega) = -K_e(\Omega)$ and $E_X(\Omega) = -K_X(\Omega) = -3\alpha_X/2m_X$. Figure 4d offers the percent ratio of the kinetic energy of proton/$\mu^+$ to the electronic kinetic energy in all the considered protonic/muonic basins, $100 \times K_X(\Omega)/K_e(\Omega)$, which for basins without proton/$\mu^+$ leakage is equal to: $100 \times E_X(\Omega)/E_e(\Omega)$. Even in the cases of the leakage, since the leakage is relatively small, $N_X(\Omega_X) \geq 0.94$, the protonic/muonic contribution in basin energy is approximately equal to the kinetic energy per particle introduced previously, $K_X(\Omega) \approx k_x$. Based on the previously discussed ab initio results, muons are contributing ~2.3 times more to the basin energies than protons and Figure 4d demonstrates the percent ratio is always larger for the muonic basins compared to their protonic congeners. However, the variations of both protonic and muonic energy contributions in all the four series are small and variations of the percent ratio mainly reflects variations of the electronic contribution in each of the series of species. On the other hand, the electronic contribution of the protonic basins is always larger than that of their muonic congeners and this is also in line with previously stated result that electrons circulating massive particles have larger kinetic energies than those circulating lighter particles. Overall, the electronic contribution dominants, and the basin energy of the protonic basins are always more negative than their muonic congeners.

Basin properties may be used to portray a more detailed picture of bonding in the protonic/muonic species complementing the primary image emerged from the topological analysis. The computed atomic charges conform to the view that protonic/muonic species containing electropositive central atoms from the extreme left-hand side of the rows are ionic systems. This is also in line with computed small electronic delocalization index ($DI^e$) for these species, which has been introduced recently in addition to the $LI^e$ within the context of the MC-QTAIM.[36]  However, for species containing electronegative elements from the extreme right-hand side of the rows, the computed atomic charges and the $DI^e$ unravel covalent bonding character with simultaneously appreciable ionic contribution.  The simultaneous covalent-ionic character strengths the bonding of protonic/muonic basins to their neighbors in line with the well-known orbitals based models of bonding.[56]  Once again, in line with the results of the topological analysis, the quantitative differences are subtle and no qualitative dissimilarity seems to be present between the mechanism of bonding in the protonic (hydrides) and the congener muonic species.

## 5. Conclusion and Prospects

In a comment on our previous computational paper on the MC-QTAIM analysis of selected muonic hydrogen species,[30] Philip Ball in his Crucible column in the Chemistry World proposed that the results point to the fact $\mu^+$ may be placed as a new member in the box of hydrogen in the PT.[64]  Although the original study was on shaky grounds with the limited considered species, present study seems to affirm that $\mu^+$ not only from kinetics point of view,[13,14] but also from the structural viewpoint is a lighter isotope of hydrogen.  Accordingly, the similarities of $\mu^+$ and proton in their bonding modes with a relatively diverse set of elements further justifies placing $\mu^+$ in the box of hydrogen.  The same reasoning may be applied to place the muonic Helium, as a composite system, also in the box of hydrogen.  Probably, one must accept at this stage that the traditional viewpoint of an element in microscopic level, as an atom composed of a certain number of electrons and a nucleus which is composed mainly of protons and neutrons, is too restrictive to be applied to the "exotic chemistry".  Although speculative, but one may imagine that if both types of muons were abundant and had a sufficiently long life-times, they would be traced as chemical elements based on their distinct structural and kinetics properties, and the emerging concept of chemical element in the course of history of chemistry would not be tied solely to the usual nuclei and the number of protons in nucleus. In a broader perspective, the basic idea of searching for the AIM structure of various exotic species, to be used as a probe to identify new potential chemical elements, is a novel possibility that its implications are quite beyond the case of muons and the muonic species.

On the other hand, recent interesting studies demonstrate that $\mu^+$, in contrast to its similarities to proton, may also yield unique modes of bonding that are peculiar to the heavier isotopes of hydrogen.[65-67] Such observations point to the interesting possibility that the MC-QTAIM analysis of the muonic species is not inevitably always similar to the protonic species and novel muonic bonding modes may also emerge from future MC-QTAIM studies.

Additionally, recent interest in various muon spin related spectroscopes and their applications in chemistry trigger the need for structural elucidation of the muonic species that materialize from exposing usual molecules to muonic beams produced in the particle accelartors.[4] Tracing $\mu^+$ trapping sites of a host molecule is a key step to interpret the results of muon spin related spectroscopies,[4,17-29] and various theoretical and computational strategies have been developed for this propose.[68-71] The extended NEO code may also serve for this purpose since it is competent to be used for optimizing the pseudo-geometries of the muonic molecules as demonstrated in present study while searching for transition states opens the door for studying elementary chemical steps in $\mu^+$ involved chemical reactions. Advanced multi-configurational post-NEO-HF methodologies are now under consideration in our lab to be used for studying muonic species beyond the NEO-HF level.

# Acknowledgments

The authors are grateful to Cina Foroutan-Nejad, Rohoullah Firouzi and Masumeh Gharabaghi for their detailed reading of a previous draft of this paper and helpful suggestions.

**Figure Captions**

**Figure- 1** (a) The difference of the mean inter-nuclear distances of each pair of the congener protonic and muonic species. (b) The kinetic energy per particle for protons in the protonic species (blue, dashed) and for muons in the muonic species (red, line). (c) The graph of the kinetic energy per particle for protons versus the optimized exponents for the protonic species. The blue line is the equation $k_H = 3\alpha_H/2m_H$ where $m_H = 1836$ has been used. (d) The graph of the kinetic energy per particle for muons versus the optimized exponents for the muonic species. The blue line is the equation $k_\mu = 3\alpha_\mu/2m_\mu$ where $m_\mu = 206$ has been used. Throughout all panels the species containing central atoms from the second and third rows of the PT have been distinguished with squares and circles, respectively.

**Figure- 2** The MGs of (a) $Li\mu$, (b) $Be\mu_2$, (c) $B\mu_3$, (d) $C\mu_4$, (e) $N\mu_3$, (f) ($O^{2-}, 2\mu^+$) and (g) ($F^-, \mu^+$). The purple and green spheres are the (3,-3) CPs and LCPs, respectively. The (3, -3) CPs at (or very near) to the central nuclei have not been depicted for clarity of the shape. Each white line is one of the gradient paths on the inter-atomic surface and the blue spherical mesh is the iso-density surface of the muonic one-particle density, $\rho_\mu = 10^{-3}$ au. It must be noted that the depicted gradient paths of $C\mu_4$ species, in contrast to the other species, do not lay in the same plan and are not crossing (For a more detailed 3D view of the MG and inter-atomic surfaces of $C\mu_4$ see Figure 3).

**Figure- 3** The MG and two selected inter-atomic surfaces of $C\mu_4$ molecule. The purple and green spheres are the (3,-3) CPs and LCPs, respectively while the white lines are the gradients all laying on the inter-atomic surfaces delineating two muonic basins from the central carbon atom. Figure S2 in supporting information depicts the MGs and inter-atomic surfaces of all the muonic species from the second row of the PT

**Figure- 4** (a) The atomic charge of the protonic (blue, dashed) and the muonic (red, line) basins. (b) The percent localization of the protonic (blue, dashed) and the muonic (red, line) basins. (c) The difference between molecular volumes of each pair of the congener muonic and protonic species (dashed) and the difference between atomic volumes (line) of the muonic and the protonic basins in each pair of the congener muonic and protonic species. (d) The percentage ratio of the protonic/muonic contribution of the kinetic energy to the electronic kinetic energy in the protonic (blue, dashed) and the muonic (red, line) basins. Throughout all panels species containing central atoms from the second and third rows of the PT have been distinguished with squares and circles, respectively.

Table 1- The results of the ab initio calculations at NEO-HF/[6-311+g(d)/4s1p:1s] computational level on both protonic and muonic species. The symbols "M" and "X" stand for the central atoms and protons/muons in the considered species, respectively. "M-X" and "X-M-X" are the mean internuclear distances and angles, respectively, "$K_X$" and "$K_e$" are the total kinetic energy of protons/muons and the electronic kinetic energy, respectively, "$\alpha_X$" is the optimized exponents of the s-type Gaussian functions used to represent the protonic/muonic orbitals. All results are offered in atomic units.

**Proton**

|    | M-X   | X-M-X  | Energy    | $K_X$  | $K_e$    | virial ratio | $\alpha_X$ |
|----|-------|--------|-----------|--------|----------|--------------|------------|
| Li | 3.095 | --     | -7.9495   | 0.0165 | 7.9276   | 2.0007       | 20.2       |
| Be | 2.569 | 180    | -15.6919  | 0.0358 | 15.6519  | 2.0003       | 21.9       |
| B  | 2.295 | 120    | -26.2733  | 0.0567 | 26.2089  | 2.0003       | 23.1       |
| C  | 2.092 | 109.47 | -40.0409  | 0.0768 | 39.9514  | 2.0003       | 23.5       |
| N  | 1.933 | 108.81 | -56.0883  | 0.0575 | 56.0122  | 2.0003       | 23.5       |
| O  | 1.820 | 106.45 | -75.9705  | 0.0377 | 75.9220  | 2.0001       | 23.1       |
| F  | 1.737 | --     | -100.0132 | 0.0183 | 99.9923  | 2.0000       | 22.4       |
| Na | 3.677 | --     | -162.3438 | 0.0159 | 162.1981 | 2.0008       | 19.5       |
| Mg | 3.283 | 180    | -200.6564 | 0.0337 | 200.5002 | 2.0006       | 20.6       |
| Al | 3.041 | 120    | -243.5216 | 0.0530 | 243.4139 | 2.0002       | 21.6       |
| Si | 2.846 | 109.47 | -291.0942 | 0.0724 | 290.9779 | 2.0002       | 22.2       |
| P  | 2.715 | 95.43  | -342.3595 | 0.0540 | 342.2915 | 2.0000       | 22.0       |
| S  | 2.568 | 93.99  | -398.6246 | 0.0353 | 398.5798 | 2.0000       | 21.6       |
| Cl | 2.452 | --     | -460.0583 | 0.0171 | 460.0589 | 2.0000       | 20.9       |

$\mu^+$

|    | M-X   | X-M-X  | Energy    | $K_X$  | $K_e$    | virial ratio | $\alpha_X$ |
|----|-------|--------|-----------|--------|----------|--------------|------------|
| Li | 3.207 | --     | -7.8919   | 0.0385 | 7.8485   | 2.0006       | 5.3        |
| Be | 2.674 | 180    | -15.5661  | 0.0845 | 15.4777  | 2.0002       | 5.8        |
| B  | 2.394 | 120    | -26.0742  | 0.1339 | 25.9333  | 2.0003       | 6.1        |
| C  | 2.183 | 109.47 | -39.7716  | 0.1810 | 39.5763  | 2.0004       | 6.2        |
| N  | 2.019 | 109.05 | -55.8869  | 0.1355 | 55.7350  | 2.0003       | 6.2        |
| O  | 1.902 | 107.56 | -75.8379  | 0.0891 | 75.7430  | 2.0001       | 6.1        |
| F  | 1.821 | --     | -99.9493  | 0.0428 | 99.9065  | 2.0000       | 5.9        |
| Na | 3.779 | --     | -162.2883 | 0.0371 | 162.1214 | 2.0008       | 5.1        |
| Mg | 3.389 | 180    | -200.5384 | 0.0791 | 200.3373 | 2.0006       | 5.4        |
| Al | 3.149 | 120    | -243.3360 | 0.1246 | 243.1566 | 2.0002       | 5.7        |
| Si | 2.951 | 109.47 | -290.8404 | 0.1701 | 290.6295 | 2.0001       | 5.8        |
| P  | 2.819 | 95.08  | -342.1707 | 0.1262 | 342.0329 | 2.0000       | 5.8        |
| S  | 2.666 | 93.93  | -398.5014 | 0.0822 | 398.4121 | 2.0000       | 5.6        |
| Cl | 2.549 | --     | -459.9989 | 0.0394 | 459.9776 | 2.0000       | 5.4        |

Table 2- Selected topological indices computed at the LCPs derived from the topological analysis performed using NEO-HF/[6-311+g(d)/4s1p:1s] wavefunction. The symbol "Γ" stands for the Gamma density, "Lap. Γ" for the Laplacian of the Gamma density, "G" for the combined Lagrangian kinetic energy density, "H" for the combined Hamiltonian energy density (see text for details). All results are offered in atomic units.

**Proton**

|      | Γ     | Lap. Γ  | G     | H      | G/Γ   |
|------|-------|---------|-------|--------|-------|
| Li   | 0.037 | 0.142   | 0.037 | -0.001 | 0.995 |
| Be   | 0.093 | 0.139   | 0.084 | -0.049 | 0.897 |
| B    | 0.176 | -0.276  | 0.119 | -0.188 | 0.676 |
| C    | 0.269 | -1.043  | 0.036 | -0.297 | 0.135 |
| N    | 0.320 | -1.692  | 0.046 | -0.469 | 0.144 |
| O    | 0.341 | -3.353  | 0.144 | -0.982 | 0.422 |
| F    | --    | --      | --    | --     | --    |
|      |       |         |       |        |       |
| Na   | 0.030 | 0.113   | 0.028 | 0.000  | 0.928 |
| Mg   | 0.051 | 0.199   | 0.055 | -0.005 | 1.076 |
| Al   | 0.079 | 0.244   | 0.088 | -0.027 | 1.112 |
| Si   | 0.116 | 0.217   | 0.128 | -0.074 | 1.106 |
| P    | 0.158 | -0.021  | 0.149 | -0.155 | 0.944 |
| S    | 0.208 | -0.620  | 0.037 | -0.192 | 0.178 |
| Cl   | 0.230 | -0.795  | 0.037 | -0.236 | 0.160 |

$\mu^+$

|      | Γ     | Lap. Γ  | G     | H      | G/Γ   |
|------|-------|---------|-------|--------|-------|
| Li   | 0.034 | 0.120   | 0.031 | -0.001 | 0.933 |
| Be   | 0.086 | 0.091   | 0.068 | -0.046 | 0.792 |
| B    | 0.166 | -0.385  | 0.077 | -0.174 | 0.467 |
| C    | 0.232 | -0.872  | 0.031 | -0.249 | 0.133 |
| N    | 0.262 | -1.646  | 0.107 | -0.519 | 0.410 |
| O    | --    | --      | --    | --     | --    |
| F    | --    | --      | --    | --     | --    |
|      |       |         |       |        |       |
| Na   | 0.028 | 0.097   | 0.024 | 0.000  | 0.868 |
| Mg   | 0.047 | 0.164   | 0.046 | -0.005 | 0.986 |
| Al   | 0.074 | 0.188   | 0.074 | -0.027 | 0.999 |
| Si   | 0.109 | 0.130   | 0.105 | -0.073 | 0.961 |
| P    | 0.150 | -0.196  | 0.101 | -0.150 | 0.672 |
| S    | 0.180 | -0.514  | 0.024 | -0.153 | 0.136 |
| Cl   | 0.189 | -0.679  | 0.051 | -0.221 | 0.271 |

Table 3- Selected topological indices computed at the (3, -3) CPs in the hydrogen/$\mu^+$ atomic basins derived from the topological analysis performed using NEO-HF/[6-311+g(d)/4s1p:1s] wavefunction. The symbol "Γ" stands for the Gamma density, "Lap. Γ" for the Laplacian of the Gamma density, "G" for the combined Lagrangian kinetic energy density, "M-LCP" and "LCP-X" for the lengths of the line paths linking each of (3, -3) CPs located in the M (central atoms) and X (proton and $\mu^+$) basins to the LCP, respectively, and "TF" for the topological floppiness index (see text for details). All results are offered in atomic units.

**Proton**

|     | Γ     | Lap. Γ   | G     | M-LCP | LCP-X | TF    |
|-----|-------|----------|-------|-------|-------|-------|
| Li  | 0.262 | -10.850  | 0.002 | 1.366 | 1.721 | 0.142 |
| Be  | 0.309 | -13.452  | 0.013 | 1.101 | 1.447 | 0.302 |
| B   | 0.344 | -15.244  | 0.035 | 0.983 | 1.281 | 0.513 |
| C   | 0.356 | -15.311  | 0.070 | 1.356 | 0.693 | 0.754 |
| N   | 0.357 | -14.177  | 0.122 | 1.496 | 0.380 | 0.896 |
| O   | 0.348 | -11.057  | 0.208 | 1.557 | 0.179 | 0.980 |
| F   | --    | --       | --    | --    | --    | --    |
| Na  | 0.244 | -9.845   | 0.001 | 1.918 | 1.753 | 0.123 |
| Mg  | 0.273 | -11.401  | 0.005 | 1.664 | 1.606 | 0.186 |
| Al  | 0.299 | -12.804  | 0.011 | 1.486 | 1.535 | 0.264 |
| Si  | 0.315 | -13.601  | 0.020 | 1.357 | 1.464 | 0.368 |
| P   | 0.312 | -13.302  | 0.030 | 1.291 | 1.397 | 0.507 |
| S   | 0.302 | -12.511  | 0.045 | 1.693 | 0.838 | 0.689 |
| Cl  | 0.286 | -11.395  | 0.062 | 1.844 | 0.561 | 0.802 |

**$\mu^+$**

|     | Γ     | Lap. Γ  | G     | M-LCP | LCP-X | TF    |
|-----|-------|---------|-------|-------|-------|-------|
| Li  | 0.191 | -3.683  | 0.001 | 1.395 | 1.791 | 0.177 |
| Be  | 0.228 | -4.594  | 0.007 | 1.128 | 1.496 | 0.378 |
| B   | 0.254 | -5.107  | 0.018 | 1.040 | 1.280 | 0.651 |
| C   | 0.262 | -4.703  | 0.042 | 1.554 | 0.520 | 0.883 |
| N   | 0.266 | -3.911  | 0.082 | 1.658 | 0.195 | 0.987 |
| O   | --    | --      | --    | --    | --    | --    |
| F   | --    | --      | --    | --    | --    | --    |
| Na  | 0.178 | -3.336  | 0.000 | 1.950 | 1.814 | 0.155 |
| Mg  | 0.201 | -3.920  | 0.002 | 1.692 | 1.665 | 0.234 |
| Al  | 0.220 | -4.392  | 0.005 | 1.514 | 1.591 | 0.335 |
| Si  | 0.231 | -4.608  | 0.010 | 1.387 | 1.505 | 0.472 |
| P   | 0.227 | -4.381  | 0.015 | 1.357 | 1.396 | 0.661 |
| S   | 0.219 | -3.979  | 0.025 | 1.947 | 0.628 | 0.825 |
| Cl  | 0.204 | -3.372  | 0.037 | 2.038 | 0.387 | 0.926 |

Table 4- The electronic contribution of selected topological indices computed at the LCPs and the (3, -3) CPs in the hydrogen/$\mu^+$ atomic basins derived from the topological analysis performed using NEO-HF/[6-311+g(d)/4s1p:1s] wavefunction. The symbol "$\rho_e$" stands for the one-particle electron density, "Lap. $\rho_e$" for the Laplacian of the one-particle electron density, "$G_e$" for the electronic Lagrangian kinetic energy density, and "$H_e$" for the electronic Hamiltonian energy density (see text for details). All results are offered in atomic units.

**Proton**

| | LCP | | | | | (3, -3) | | |
|---|---|---|---|---|---|---|---|---|
| | $\rho_e$ | Lap. $\rho_e$ | $G_e$ | $H_e$ | $G_e/\rho_e$ | $\rho_e$ | Lap. $\rho_e$ | $G_e$ |
| Li | 0.037 | 0.142 | 0.037 | -0.001 | 0.995 | 0.237 | -4.817 | 0.000 |
| Be | 0.093 | 0.139 | 0.084 | -0.049 | 0.897 | 0.281 | -6.219 | 0.001 |
| B  | 0.176 | -0.276 | 0.119 | -0.188 | 0.676 | 0.314 | -7.334 | 0.004 |
| C  | 0.269 | -1.043 | 0.036 | -0.297 | 0.135 | 0.327 | -7.601 | 0.012 |
| N  | 0.320 | -1.697 | 0.045 | -0.469 | 0.141 | 0.330 | -7.379 | 0.025 |
| O  | 0.340 | -3.747 | 0.051 | -0.988 | 0.150 | 0.326 | -6.270 | 0.041 |
| F  | -- | -- | -- | -- | -- | -- | -- | -- |
| Na | 0.030 | 0.113 | 0.028 | 0.000 | 0.928 | 0.221 | -4.301 | 0.000 |
| Mg | 0.051 | 0.199 | 0.055 | -0.005 | 1.076 | 0.248 | -5.064 | 0.000 |
| Al | 0.079 | 0.244 | 0.088 | -0.027 | 1.112 | 0.272 | -5.791 | 0.001 |
| Si | 0.116 | 0.217 | 0.128 | -0.074 | 1.106 | 0.287 | -6.280 | 0.002 |
| P  | 0.158 | -0.021 | 0.149 | -0.155 | 0.944 | 0.285 | -6.226 | 0.005 |
| S  | 0.208 | -0.620 | 0.037 | -0.192 | 0.178 | 0.276 | -6.006 | 0.010 |
| Cl | 0.230 | -0.795 | 0.037 | -0.236 | 0.160 | 0.262 | -5.701 | 0.015 |

**$\mu^+$**

| | LCP | | | | | (3, -3) | | |
|---|---|---|---|---|---|---|---|---|
| | $\rho_e$ | Lap. $\rho_e$ | $G_e$ | $H_e$ | $G_e/\rho_e$ | $\rho_e$ | Lap. $\rho_e$ | $G_e$ |
| Li | 0.034 | 0.120 | 0.031 | -0.001 | 0.933 | 0.161 | -1.792 | 0.000 |
| Be | 0.086 | 0.091 | 0.068 | -0.046 | 0.792 | 0.195 | -2.306 | 0.001 |
| B  | 0.166 | -0.385 | 0.077 | -0.174 | 0.467 | 0.219 | -2.648 | 0.004 |
| C  | 0.231 | -0.919 | 0.022 | -0.252 | 0.096 | 0.229 | -2.486 | 0.012 |
| N  | 0.255 | -1.691 | 0.032 | -0.454 | 0.124 | 0.239 | -2.360 | 0.025 |
| O  | -- | -- | -- | -- | -- | -- | -- | -- |
| F  | -- | -- | -- | -- | -- | -- | -- | -- |
| Na | 0.028 | 0.097 | 0.024 | 0.000 | 0.868 | 0.149 | -1.612 | 0.000 |
| Mg | 0.047 | 0.164 | 0.046 | -0.005 | 0.986 | 0.170 | -1.923 | 0.000 |
| Al | 0.074 | 0.188 | 0.074 | -0.027 | 0.999 | 0.188 | -2.179 | 0.001 |
| Si | 0.109 | 0.130 | 0.105 | -0.073 | 0.961 | 0.198 | -2.328 | 0.002 |
| P  | 0.150 | -0.196 | 0.101 | -0.150 | 0.672 | 0.195 | -2.226 | 0.005 |
| S  | 0.180 | -0.532 | 0.021 | -0.154 | 0.117 | 0.189 | -2.076 | 0.009 |
| Cl | 0.188 | -0.784 | 0.023 | -0.219 | 0.123 | 0.178 | -1.851 | 0.013 |

Table 5- Selected results of the basin integrations of the protonic species performed using NEO-HF/[6-311+g(d)/4s1p:1s] wavefunction. The symbols "*LI*" and "*DI*" stand for the electronic localization and delocalization indices within and between the M (central atom) and X (proton and $\mu^+$) atomic basins, respectively, while the "electric dipoles" are the intra-atomic electric dipoles (see text for details). All results are offered in atomic units.

| *M-basin* | basin energy | electron population | proton population | LI | atomic volume | electric dipole | DI (M,X) | (X,X) |
|---|---|---|---|---|---|---|---|---|
| Li | -7.3634 | 2.089 | 0.000 | 1.99 | 31.2 | 0.00 | 0.20 | -- |
| Be | -14.2411 | 2.303 | 0.000 | 2.01 | 21.8 | 0.00 | 0.29 | 0.09 |
| B | -23.8279 | 2.992 | 0.000 | 2.17 | 27.8 | 0.00 | 0.55 | 0.14 |
| C | -37.7701 | 6.155 | 0.000 | 4.19 | 91.2 | 0.00 | 0.98 | 0.04 |
| N | -54.9001 | 8.407 | 0.000 | 7.23 | 154.4 | 0.01 | 0.79 | 0.01 |
| O | -75.4278 | 9.411 | 0.009 | 8.90 | 156.0 | 0.46 | 0.51 | 0.00 |
| Na | -161.8094 | 10.194 | 0.000 | 10.00 | 85.7 | 0.00 | 0.41 | -- |
| Mg | -199.4250 | 10.397 | 0.000 | 10.01 | 77.4 | 0.00 | 0.38 | 0.04 |
| Al | -241.4140 | 10.642 | 0.000 | 10.05 | 50.9 | 0.00 | 0.39 | 0.10 |
| Si | -288.0154 | 11.069 | 0.000 | 10.14 | 39.4 | 0.00 | 0.47 | 0.12 |
| P | -340.0357 | 13.243 | 0.000 | 11.99 | 149.7 | 2.29 | 0.83 | 0.13 |
| S | -397.5644 | 16.085 | 0.000 | 14.99 | 228.2 | 0.59 | 1.09 | 0.03 |
| Cl | -459.6605 | 17.390 | 0.000 | 16.94 | 244.3 | 0.20 | 0.90 | -- |

| *X-basin* | basin energy | electron population | proton population | LI | atomic volume | electric dipole |
|---|---|---|---|---|---|---|
| Li | -0.5861 | 1.912 | 1.000 | 1.81 | 201.7 | 0.46 |
| Be | -0.7254 | 1.849 | 1.000 | 1.66 | 142.7 | 0.59 |
| B | -0.8151 | 1.669 | 1.000 | 1.26 | 91.7 | 0.39 |
| C | -0.5677 | 0.961 | 1.000 | 0.41 | 51.2 | 0.16 |
| N | -0.3961 | 0.531 | 1.000 | 0.13 | 29.2 | 0.20 |
| O | -0.2713 | 0.294 | 0.991 | 0.04 | 16.8 | 0.15 |
| Na | -0.5344 | 1.806 | 1.000 | 1.58 | 198.1 | 0.18 |
| Mg | -0.6157 | 1.802 | 1.000 | 1.59 | 163.1 | 0.39 |
| Al | -0.7025 | 1.786 | 1.000 | 1.49 | 129.7 | 0.45 |
| Si | -0.7697 | 1.733 | 1.000 | 1.32 | 100.8 | 0.47 |
| P | -0.7746 | 1.586 | 1.000 | 1.04 | 81.0 | 0.36 |
| S | -0.5301 | 0.957 | 1.000 | 0.39 | 52.2 | 0.10 |
| Cl | -0.3978 | 0.610 | 1.000 | 0.16 | 35.0 | 0.15 |

Table 6- Selected results of the basin integrations of the muonic species performed using NEO-HF/[6-311+g(d)/4s1p:1s] wavefunction. The symbols "*LI*" and "*DI*" stand for the electronic localization and delocalization indices within and between the M (central atom) and X (proton and $\mu^+$) atomic basins, respectively, while the electric dipoles are the intra-atomic electric dipoles (see text for details). All results are offered in atomic units.

| *M-basin* | basin energy | electron population | $\mu^+$ population | *LI* | atomic volume | electric dipole | *DI* (M,X) | (X,X) |
|---|---|---|---|---|---|---|---|---|
| Li | -7.3627 | 2.095 | 0.000 | 1.99 | 32.6 | 0.00 | 0.21 | -- |
| Be | -14.2385 | 2.321 | 0.000 | 2.02 | 23.7 | 0.00 | 0.31 | 0.09 |
| B | -23.9208 | 3.213 | 0.000 | 2.26 | 37.8 | 0.00 | 0.64 | 0.13 |
| C | -38.0276 | 6.929 | 0.005 | 5.07 | 125.4 | 0.00 | 0.93 | 0.02 |
| N | -55.0219 | 8.818 | 0.056 | 7.87 | 178.2 | 0.29 | 0.63 | 0.01 |
| Na | -161.8087 | 10.21 | 0.000 | 10.00 | 89.7 | 0.00 | 0.41 | -- |
| Mg | -199.4221 | 10.42 | 0.000 | 10.02 | 81.8 | 0.00 | 0.40 | 0.04 |
| Al | -241.4193 | 10.70 | 0.000 | 10.07 | 55.4 | 0.00 | 0.42 | 0.11 |
| Si | -288.0629 | 11.22 | 0.000 | 10.19 | 46.8 | 0.00 | 0.51 | 0.12 |
| P | -340.1853 | 13.58 | 0.000 | 12.22 | 162.7 | 2.11 | 0.91 | 0.12 |
| S | -397.7047 | 16.52 | 0.001 | 15.54 | 247.3 | 0.01 | 0.98 | 0.02 |
| Cl | -459.7106 | 17.56 | 0.014 | 17.20 | 253.3 | 0.53 | 0.71 | -- |

| *X-basin* | basin energy | electron population | $\mu^+$ population | *LI* | atomic volume | electric dipole |
|---|---|---|---|---|---|---|
| Li | -0.5292 | 1.905 | 1.000 | 1.80 | 214.4 | 0.55 |
| Be | -0.6638 | 1.839 | 1.000 | 1.64 | 151.2 | 0.68 |
| B | -0.7178 | 1.596 | 1.000 | 1.15 | 95.4 | 0.38 |
| C | -0.4360 | 0.768 | 0.999 | 0.27 | 48.1 | 0.22 |
| N | -0.2883 | 0.394 | 0.944 | 0.07 | 25.7 | 0.20 |
| Na | -0.4795 | 1.789 | 1.000 | 1.58 | 209.6 | 0.22 |
| Mg | -0.5582 | 1.788 | 1.000 | 1.57 | 172.8 | 0.46 |
| Al | -0.6389 | 1.767 | 1.000 | 1.45 | 137.2 | 0.52 |
| Si | -0.6944 | 1.695 | 1.000 | 1.26 | 106.3 | 0.51 |
| P | -0.6618 | 1.472 | 1.000 | 0.90 | 83.2 | 0.31 |
| S | -0.3984 | 0.738 | 1.000 | 0.24 | 48.0 | 0.18 |
| Cl | -0.2883 | 0.442 | 0.986 | 0.09 | 30.5 | 0.17 |

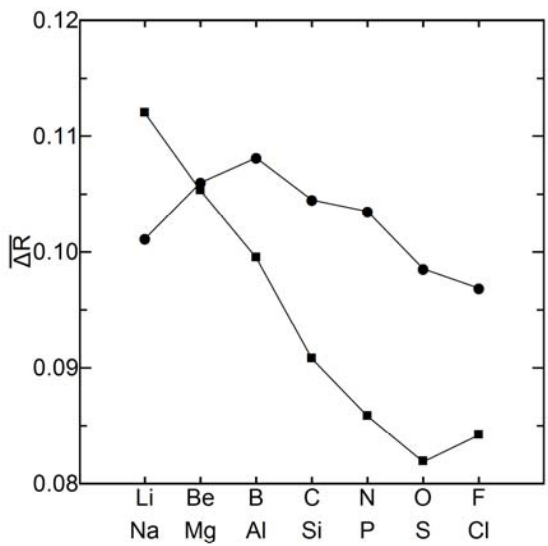

(a)

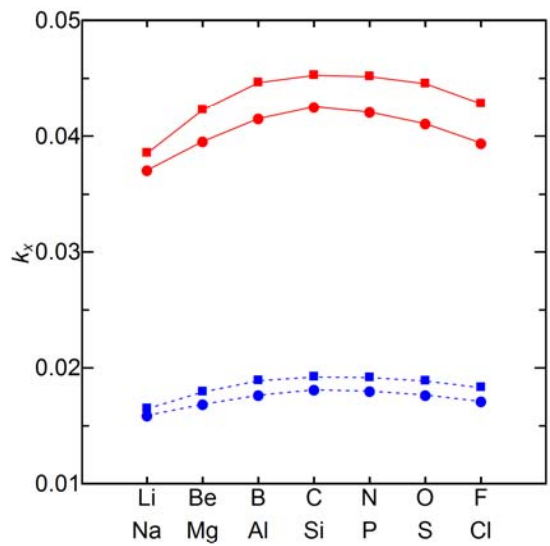

(b)

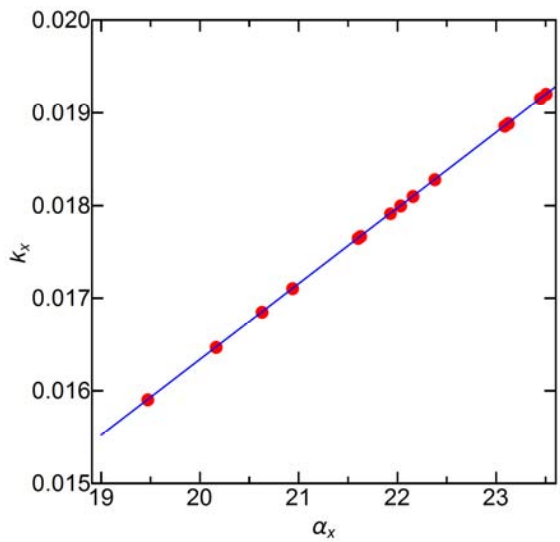

(c)

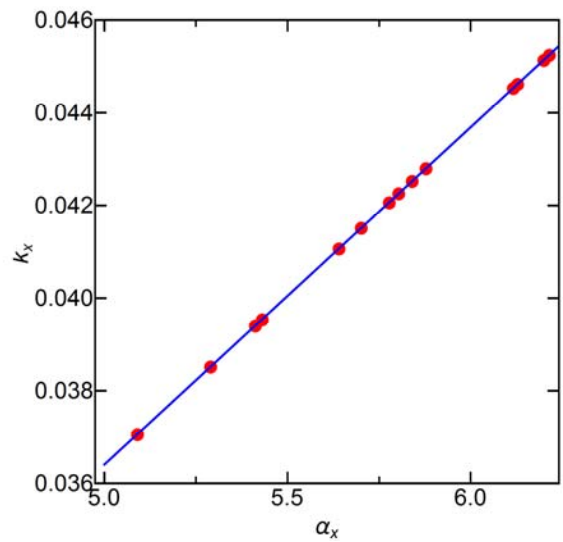

(d)

**Figure-1**

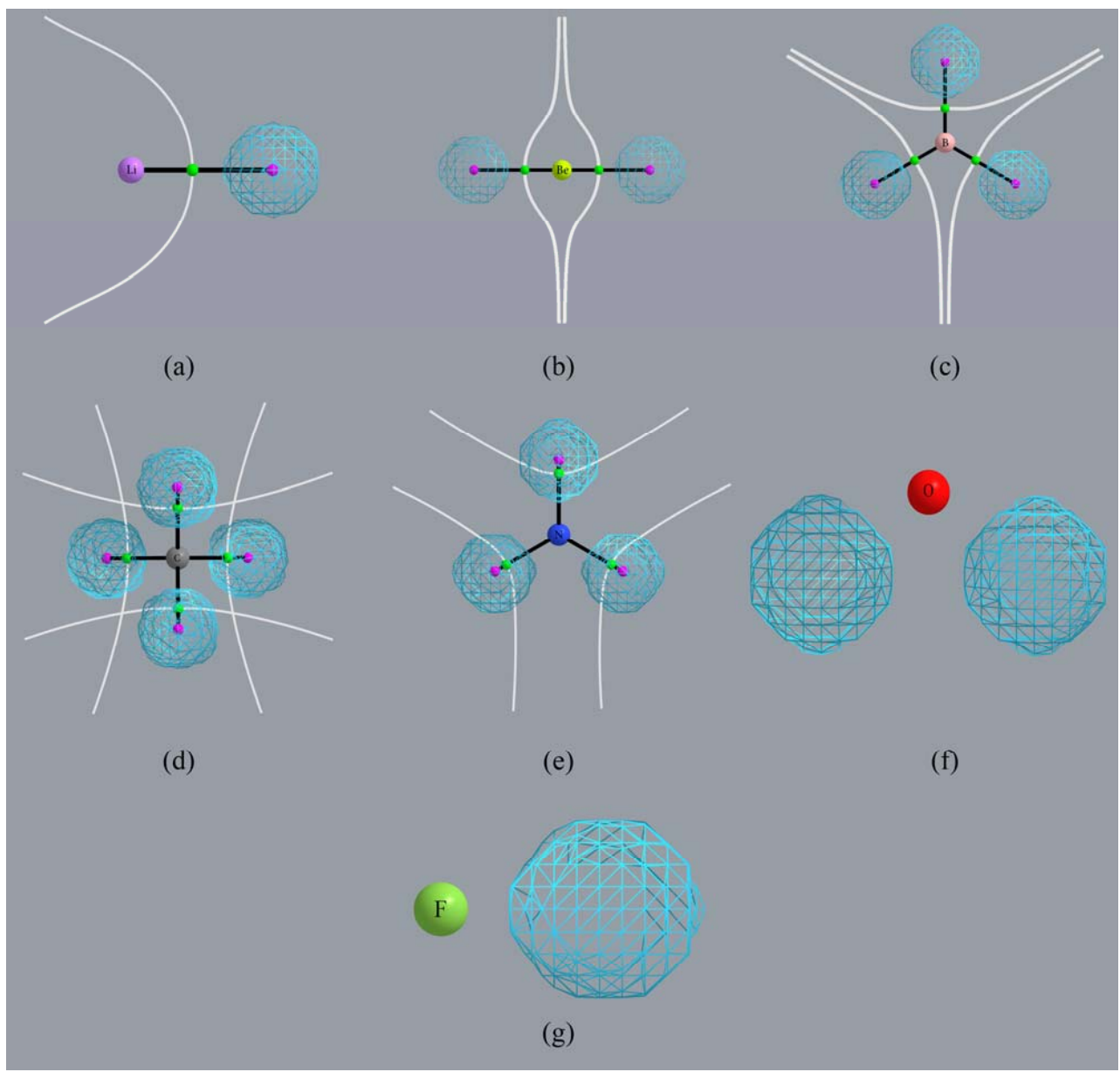

**Figure-2**

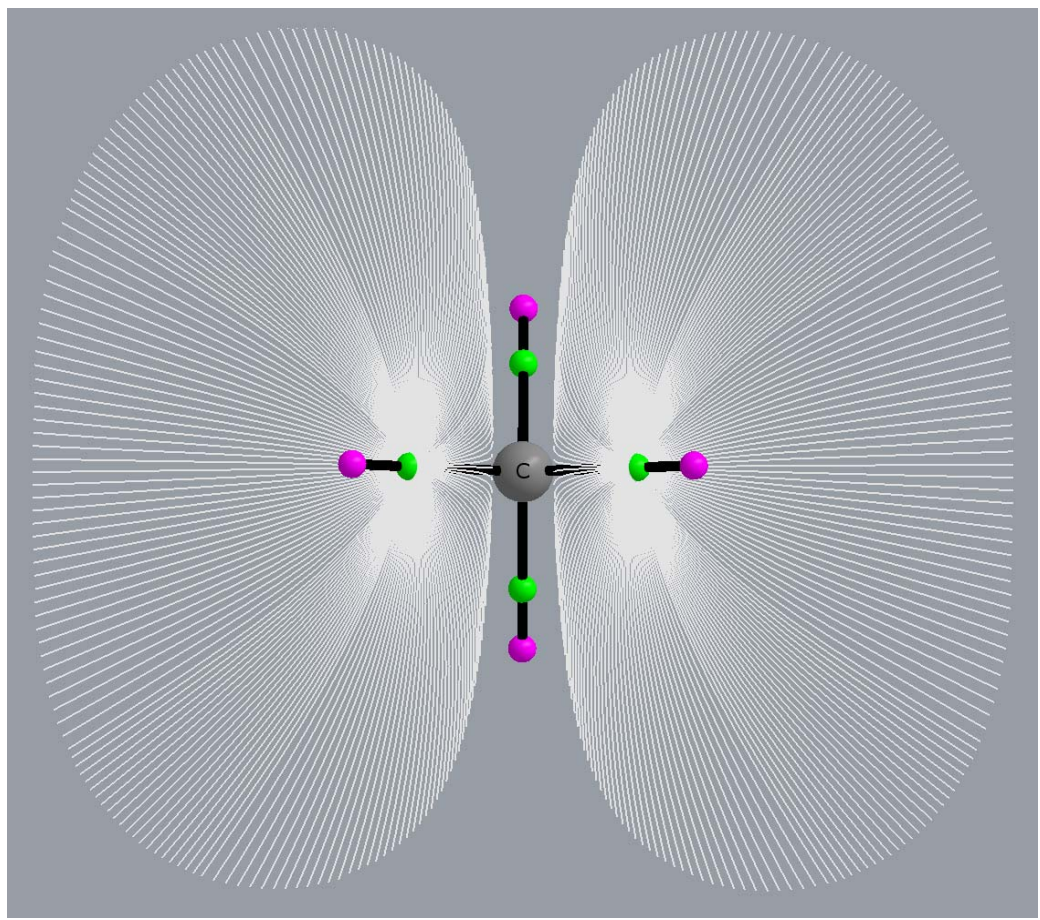

**Figure-3**

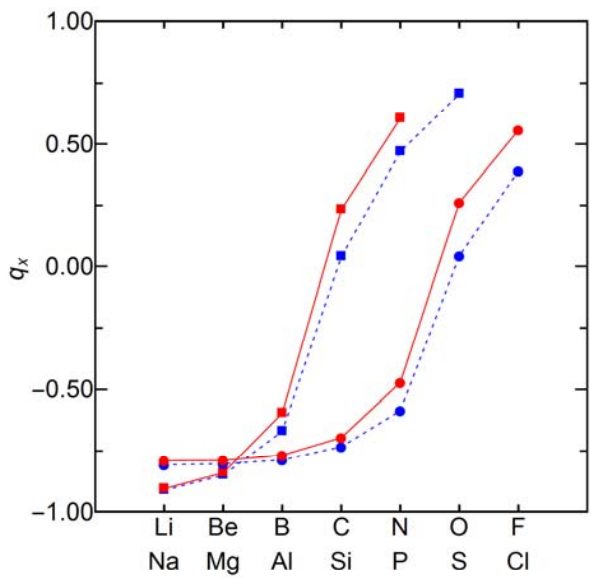

(a)

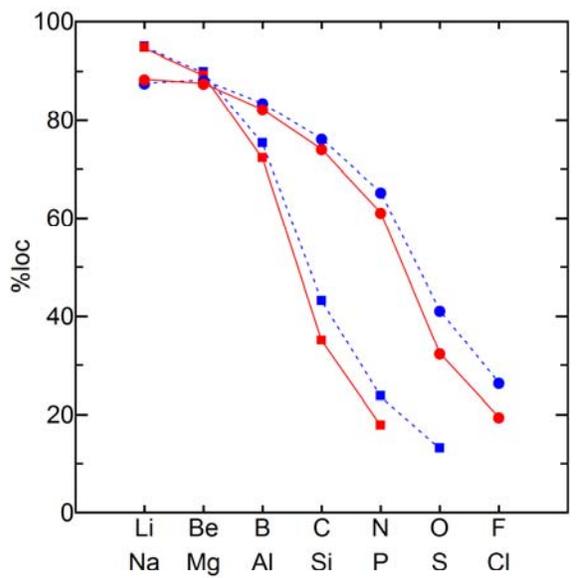

(b)

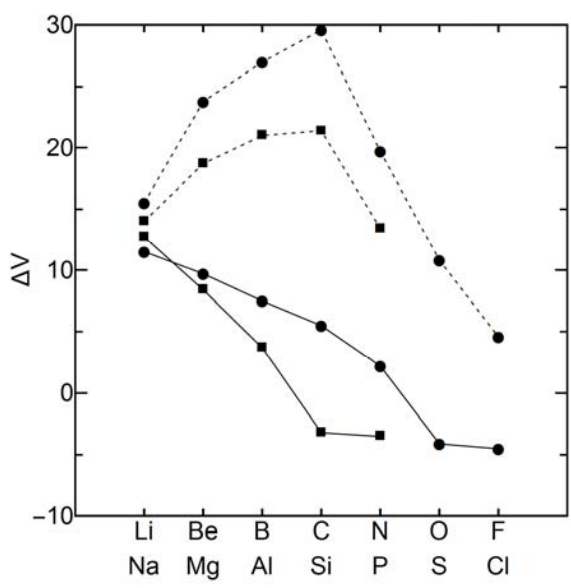

(c)

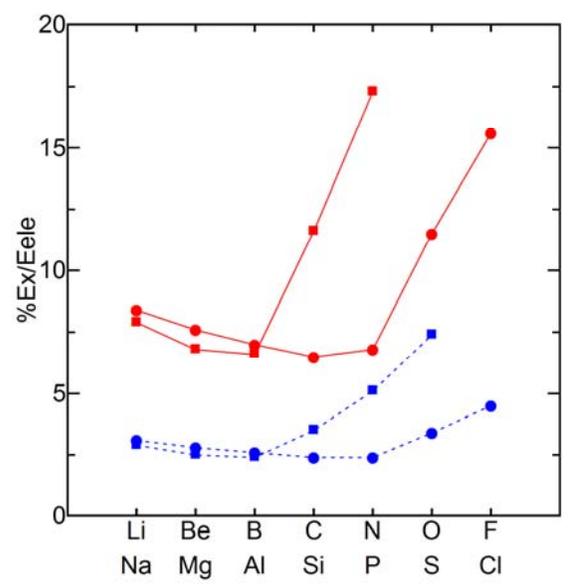

(d)

**Figure-4**

# Supporting Information

## Where to place the positive muon in the Periodic Table?

Mohammad Goli and Shant Shahbazian[*]

*Faculty of Chemistry, Shahid Beheshti University, G. C. , Evin, Tehran, Iran, 19839, P.O. Box 19395-4716.*

Tel/Fax: 98-21-22431661

E-mail:
(Shant Shahbazian) chemist_shant@yahoo.com

[*] Corresponding author

# Table of contents

Table S1- The optimized exponents of the electronic [4s1p] basis set for both protonic and muonic species.

| Proton | Li | Be | B | C | N | O | F |
|---|---|---|---|---|---|---|---|
| S | 6.863 | 8.531 | 10.158 | 10.554 | 10.490 | 7.889 | 8.161 |
| S | 1.449 | 1.884 | 2.351 | 2.487 | 2.460 | 1.633 | 1.732 |
| S | 0.372 | 0.492 | 0.630 | 0.707 | 0.717 | 0.469 | 0.534 |
| S | 0.106 | 0.134 | 0.178 | 0.213 | 0.242 | 0.140 | 0.151 |
| P | 0.345 | 0.742 | 0.981 | 1.105 | 0.772 | 0.687 | 0.829 |
| $\mu^+$ | | | | | | | |
| S | 3.537 | 4.406 | 5.165 | 3.604 | 4.016 | 4.455 | 3.858 |
| S | 0.967 | 1.310 | 1.653 | 0.894 | 1.047 | 1.214 | 1.011 |
| S | 0.286 | 0.412 | 0.533 | 0.271 | 0.340 | 0.399 | 0.350 |
| S | 0.090 | 0.121 | 0.164 | 0.068 | 0.099 | 0.121 | 0.099 |
| P | 0.308 | 0.727 | 0.878 | 0.950 | 0.691 | 0.652 | 0.787 |
| Proton | Na | Mg | Al | Si | P | S | Cl |
| S | 6.189 | 6.903 | 7.633 | 8.284 | 8.489 | 8.848 | 9.884 |
| S | 1.279 | 1.447 | 1.624 | 1.795 | 1.864 | 1.977 | 2.327 |
| S | 0.320 | 0.360 | 0.411 | 0.461 | 0.493 | 0.528 | 0.623 |
| S | 0.091 | 0.100 | 0.121 | 0.142 | 0.163 | 0.177 | 0.215 |
| P | 0.253 | 0.477 | 0.482 | 0.559 | 0.546 | 0.561 | 0.549 |
| $\mu^+$ | | | | | | | |
| S | 3.240 | 3.733 | 4.069 | 4.432 | 4.272 | 4.548 | 5.606 |
| S | 0.864 | 1.039 | 1.150 | 1.284 | 1.238 | 1.364 | 1.818 |
| S | 0.248 | 0.302 | 0.338 | 0.385 | 0.391 | 0.436 | 0.551 |
| S | 0.078 | 0.089 | 0.108 | 0.128 | 0.147 | 0.162 | 0.201 |
| P | 0.234 | 0.464 | 0.430 | 0.516 | 0.474 | 0.494 | 0.492 |

Table S2- The results of the ab initio NEO-CI/[6-311+g(d)/4s1p:1s] calculations on the muonic species. The symbol "$K_X$" stands for the total kinetic energy of muons, and "corr. Energy" is the electronic correlation energy computed as the energy difference between total energies of NEO-HF/[6-311+g(d)/4s1p:1s] and NEO-CI/[6-311+g(d)/4s1p:1s] computational levels. All results are offered in atomic units.

|  | Active space | Number of the used determinants | Energy | $K_X$ | virial ratio | corr. Energy |
|---|---|---|---|---|---|---|
| **Li** | (2,5) | 25 | -7.8920 | 0.0385 | 2.0007 | 0.0002 |
| **Be** | (4,6) | 225 | -15.5664 | 0.0845 | 2.0003 | 0.0003 |
| **B** | (6,7) | 1225 | -26.0754 | 0.1339 | 2.0006 | 0.0012 |
| **C** | (8,8) | 4900 | -39.7722 | 0.1810 | 2.0005 | 0.0006 |
| **N** | (6,6) | 400 | -55.8881 | 0.1355 | 2.0005 | 0.0012 |
| **O** | (8,8) | 4900 | -75.8512 | 0.0891 | 2.0021 | 0.0133 |
| **F** | (8,8) | 4900 | -99.9603 | 0.0428 | 2.0018 | 0.0110 |
| **Na** | (2,5) | 25 | -162.2887 | 0.0371 | 2.0008 | 0.0004 |
| **Mg** | (4,6) | 225 | -200.5385 | 0.0791 | 2.0006 | 0.0001 |
| **Al** | (6,7) | 1225 | -243.3367 | 0.1246 | 2.0002 | 0.0007 |
| **Si** | (8,8) | 4900 | -290.8410 | 0.1701 | 2.0002 | 0.0006 |
| **P** | (6,6) | 400 | -342.1709 | 0.1262 | 2.0000 | 0.0002 |
| **S** | (8,8) | 4900 | -398.5028 | 0.0822 | 2.0001 | 0.0014 |
| **Cl** | (8,8) | 4900 | -460.0010 | 0.0394 | 2.0000 | 0.0021 |

Table S3- The protonic/muonic contribution of selected topological indices computed at the LCPs derived from the topological analysis performed using NEO-HF/[6-311+g(d)/4s1p:1s] wavefunction. The symbol "$\rho_x$" stands for the one-particle protonic/muonic density, "Lap. $\rho_x$" for the Laplacian of the one-particle protonic/muonic density, "$G_x$" for the protonic/muonic Lagrangian kinetic energy density, and "$H_x$" for the protonic/muonic Hamiltonian energy density (see text for details). The mass-scaled quantities have been derived by dividing each quantity to the mass of proton/$\mu^+$. All results are offered in atomic units.

**Proton**

|    | $\rho_x$ | Lap. $\rho_x$ | mass-scaled $\rho_x$ | mass-scaled Lap. $\rho_x$ | $G_x$ | $H_x$ |
|----|----------|---------------|----------------------|---------------------------|-------|-------|
| Li | 0.000 | 0.000 | 0.000 | 0.000 | 0.000 | 0.000 |
| Be | 0.000 | 0.000 | 0.000 | 0.000 | 0.000 | 0.000 |
| B  | 0.000 | 0.000 | 0.000 | 0.000 | 0.000 | 0.000 |
| C  | 0.000 | 0.000 | 0.000 | 0.000 | 0.000 | 0.000 |
| N  | 0.007 | 10.351 | 0.000 | 0.006 | 0.001 | 0.001 |
| O  | 2.316 | 723.883 | 0.001 | 0.394 | 0.093 | 0.006 |
| F  | -- | -- | -- | -- | -- | -- |
| Na | 0.000 | 0.000 | 0.000 | 0.000 | 0.000 | 0.000 |
| Mg | 0.000 | 0.000 | 0.000 | 0.000 | 0.000 | 0.000 |
| Al | 0.000 | 0.000 | 0.000 | 0.000 | 0.000 | 0.000 |
| Si | 0.000 | 0.000 | 0.000 | 0.000 | 0.000 | 0.000 |
| P  | 0.000 | 0.000 | 0.000 | 0.000 | 0.000 | 0.000 |
| S  | 0.000 | 0.000 | 0.000 | 0.000 | 0.000 | 0.000 |
| Cl | 0.000 | 0.021 | 0.000 | 0.000 | 0.000 | 0.000 |

**$\mu^+$**

|    | $\rho_x$ | Lap. $\rho_x$ | mass-scaled $\rho_x$ | mass-scaled Lap. $\rho_x$ | $G_x$ | $H_x$ |
|----|----------|---------------|----------------------|---------------------------|-------|-------|
| Li | 0.000 | 0.000 | 0.000 | 0.000 | 0.000 | 0.000 |
| Be | 0.000 | 0.000 | 0.000 | 0.000 | 0.000 | 0.000 |
| B  | 0.000 | 0.000 | 0.000 | 0.000 | 0.000 | 0.000 |
| C  | 0.058 | 9.817 | 0.000 | 0.048 | 0.009 | 0.003 |
| N  | 1.554 | 9.176 | 0.008 | 0.045 | 0.076 | -0.065 |
| O  | -- | -- | -- | -- | -- | -- |
| F  | -- | -- | -- | -- | -- | -- |
| Na | 0.000 | 0.000 | 0.000 | 0.000 | 0.000 | 0.000 |
| Mg | 0.000 | 0.000 | 0.000 | 0.000 | 0.000 | 0.000 |
| Al | 0.000 | 0.000 | 0.000 | 0.000 | 0.000 | 0.000 |
| Si | 0.000 | 0.000 | 0.000 | 0.000 | 0.000 | 0.000 |
| P  | 0.000 | 0.000 | 0.000 | 0.000 | 0.000 | 0.000 |
| S  | 0.019 | 3.821 | 0.000 | 0.019 | 0.003 | 0.002 |
| Cl | 0.379 | 21.764 | 0.002 | 0.106 | 0.028 | -0.002 |

Table S4- The protonic/muonic contribution of selected topological indices computed at the (3, -3) CPs in the hydrogen/$\mu^+$ atomic basins derived from the topological analysis performed using NEO-HF/[6-311+g(d)/4s1p:1s] wavefunction. The symbol "$\rho_x$" stands for the one-particle protonic/muonic density, "Lap. $\rho_x$" for the Laplacian of the one-particle protonic/muonic density, "$G_x$" for the protonic/muonic Lagrangian kinetic energy density (see text for details). The mass-scaled quantities have been derived by dividing each quantity to the mass of proton/$\mu^+$. All results are offered in atomic units.

**Proton**

|     | $\rho_x$ | Lap. $\rho_x$ | mass-scaled $\rho_x$ | mass-scaled Lap. $\rho_x$ | $G_x$ |
|-----|----------|---------------|----------------------|---------------------------|-------|
| Li  | 45.865   | -11077.238    | 0.025                | -6.033                    | 0.002 |
| Be  | 51.142   | -13280.651    | 0.028                | -7.233                    | 0.012 |
| B   | 53.973   | -14523.018    | 0.029                | -7.910                    | 0.031 |
| C   | 53.189   | -14156.223    | 0.029                | -7.710                    | 0.058 |
| N   | 49.438   | -12481.559    | 0.027                | -6.798                    | 0.098 |
| O   | 40.599   | -8789.329     | 0.022                | -4.787                    | 0.167 |
| F   | --       | --            | --                   | --                        | --    |
| Na  | 43.594   | -10178.548    | 0.024                | -5.544                    | 0.001 |
| Mg  | 47.237   | -11635.055    | 0.026                | -6.337                    | 0.004 |
| Al  | 50.201   | -12876.353    | 0.027                | -7.013                    | 0.009 |
| Si  | 51.513   | -13440.669    | 0.028                | -7.321                    | 0.018 |
| P   | 50.480   | -12991.807    | 0.027                | -7.076                    | 0.024 |
| S   | 48.011   | -11944.087    | 0.026                | -6.505                    | 0.035 |
| Cl  | 44.354   | -10454.663    | 0.024                | -5.694                    | 0.047 |

**$\mu^+$**

|     | $\rho_x$ | Lap. $\rho_x$ | mass-scaled $\rho_x$ | mass-scaled Lap. $\rho_x$ | $G_x$ |
|-----|----------|---------------|----------------------|---------------------------|-------|
| Li  | 6.154    | -389.551      | 0.030                | -1.891                    | 0.001 |
| Be  | 6.900    | -471.295      | 0.033                | -2.288                    | 0.006 |
| B   | 7.210    | -506.642      | 0.035                | -2.459                    | 0.014 |
| C   | 6.792    | -456.767      | 0.033                | -2.217                    | 0.030 |
| N   | 5.566    | -319.477      | 0.027                | -1.551                    | 0.057 |
| O   | --       | --            | --                   | --                        | --    |
| F   | --       | --            | --                   | --                        | --    |
| Na  | 5.822    | -355.201      | 0.028                | -1.724                    | 0.000 |
| Mg  | 6.358    | -411.323      | 0.031                | -1.997                    | 0.002 |
| Al  | 6.764    | -455.903      | 0.033                | -2.213                    | 0.004 |
| Si  | 6.887    | -469.778      | 0.033                | -2.280                    | 0.008 |
| P   | 6.659    | -443.918      | 0.032                | -2.155                    | 0.011 |
| S   | 6.184    | -391.846      | 0.030                | -1.902                    | 0.016 |
| Cl  | 5.420    | -313.201      | 0.026                | -1.520                    | 0.024 |

Table S5- Selected topological indices computed at the LCPs and the (3, -3) CPs in $\mu^+$ atomic basins derived from the topological analysis using NEO-CI/[6-311+g(d)/4s1p:1s] wavefunction. The symbol "Γ" stands for the Gamma density, "Lap. Γ" for the Laplacian of the Gamma density, "G" for the combined Lagrangian kinetic energy density, "H" for the combined Hamiltonian energy density, "M-LCP" and "LCP-X" for the lengths of the line paths linking each of (3, -3) CPs located in the M (central atoms) and X basins to the LCP, respectively, and "TF" for the topological floppiness index (see text for details). All results are offered in atomic units.

|    | LCP   |        |       |        |       | (3, -3) |        |       |       |       |       |
|----|-------|--------|-------|--------|-------|---------|--------|-------|-------|-------|-------|
|    | Γ     | Lap. Γ | G     | H      | G/Γ   | Γ       | Lap. Γ | G     | M-LCP | LCP-X | TF    |
| Li | 0.034 | 0.120  | 0.031 | -0.001 | 0.934 | 0.191   | -3.678 | 0.001 | 1.396 | 1.791 | 0.177 |
| Be | 0.086 | 0.091  | 0.068 | -0.046 | 0.792 | 0.228   | -4.593 | 0.007 | 1.128 | 1.496 | 0.378 |
| B  | 0.165 | -0.384 | 0.077 | -0.173 | 0.467 | 0.254   | -5.106 | 0.018 | 1.040 | 1.280 | 0.650 |
| C  | 0.231 | -0.871 | 0.031 | -0.248 | 0.133 | 0.262   | -4.702 | 0.042 | 1.554 | 0.520 | 0.883 |
| N  | 0.262 | -1.640 | 0.107 | -0.517 | 0.410 | 0.266   | -3.915 | 0.082 | 1.657 | 0.196 | 0.987 |
| O  | --    | --     | --    | --     | --    | --      | --     | --    | --    | --    | --    |
| F  | --    | --     | --    | --     | --    | --      | --     | --    | --    | --    | --    |
| Na | 0.027 | 0.096  | 0.024 | 0.000  | 0.866 | 0.177   | -3.326 | 0.000 | 1.951 | 1.812 | 0.155 |
| Mg | 0.047 | 0.164  | 0.046 | -0.005 | 0.986 | 0.201   | -3.919 | 0.002 | 1.693 | 1.664 | 0.234 |
| Al | 0.074 | 0.188  | 0.074 | -0.027 | 0.999 | 0.220   | -4.391 | 0.005 | 1.514 | 1.590 | 0.335 |
| Si | 0.109 | 0.130  | 0.105 | -0.073 | 0.961 | 0.231   | -4.607 | 0.010 | 1.387 | 1.505 | 0.472 |
| P  | 0.150 | -0.196 | 0.101 | -0.150 | 0.671 | 0.227   | -4.380 | 0.015 | 1.357 | 1.396 | 0.661 |
| S  | 0.180 | -0.513 | 0.024 | -0.153 | 0.136 | 0.218   | -3.979 | 0.025 | 1.947 | 0.629 | 0.824 |
| Cl | 0.189 | -0.675 | 0.051 | -0.220 | 0.270 | 0.204   | -3.376 | 0.037 | 2.037 | 0.389 | 0.925 |

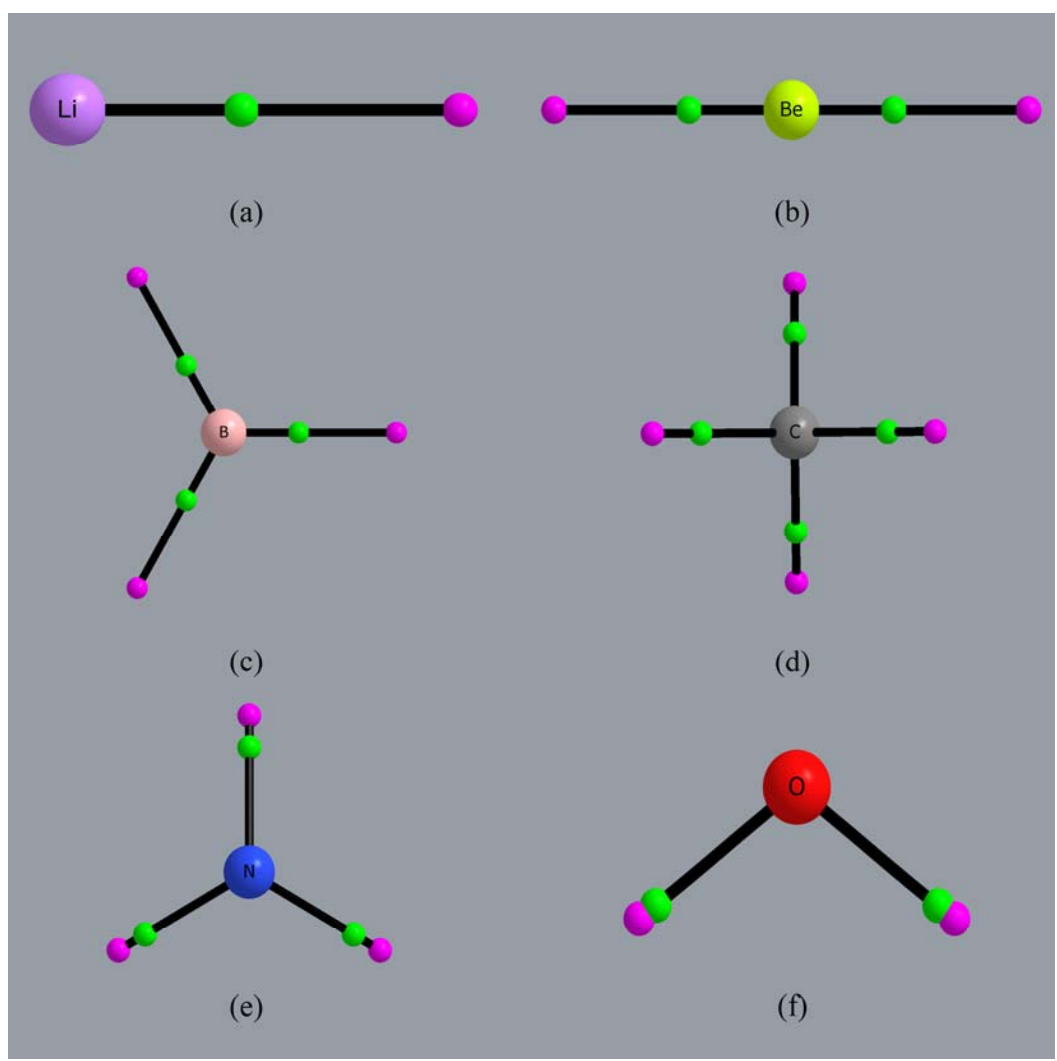

Figure S1- The MGs of the protonic species: a) LiH, (b) BeH$_2$, (c) BH$_3$, (d) CH$_4$, (e) NH$_3$ and (f) H$_2$O. The FH molecule has just a single (3, -3) CP very near to the fluorine nucleus thus it has no MG at the considered computational level. The purple and green spheres are the (3,-3) CPs and LCPs, respectively, while the black line are the line paths, i.e. gradient paths, connection the (3, -3) CPs and LCPs. The (3, -3) CPs at (or very near) to the central nuclei have not been depicted for clarity of the shapes. Except from HCl molecule that has a MG similar to LiH molecule, the congener species containing central atoms from the third row of the PT have exactly the same MG depicted in this figure.

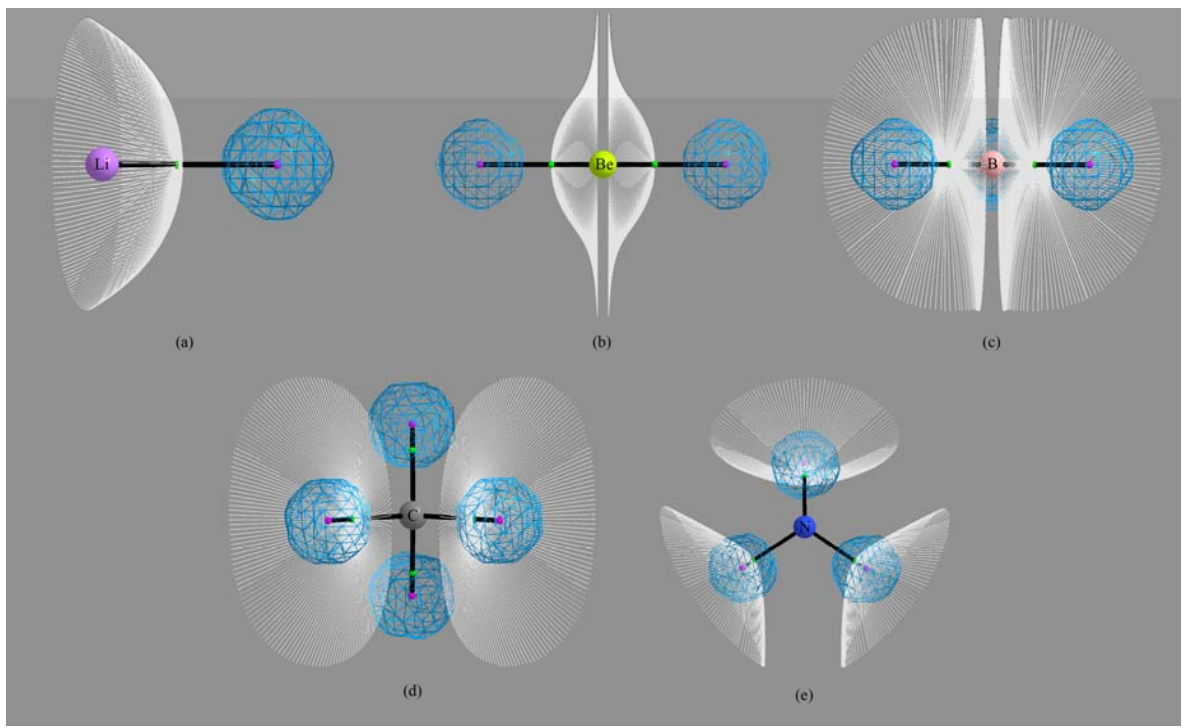

Figure S2- The MGs of the muonic species: (a) $Li\mu$, (b) $Be\mu_2$, (c) $B\mu_3$, (d) $C\mu_4$, (e) $N\mu_3$. The purple and green spheres are the (3,-3) CPs and LCPs, respectively. The ($F^-,\mu^+$) and ($O^{2-},2\mu^+$) species have just a single (3, -3) CP very near to the fluorine and oxygen nuclei thus they have no MG at the considered computational level. The purple and green spheres are the (3,-3) CPs and LCPs, respectively, while each white line is one of the gradient paths on the inter-atomic surface. The blue spherical mesh is an iso-density surface of the muonic one-particle density, $\rho_\mu(\vec{q}) = 10^{-3}\,au$.